\documentclass[10pt,aps,prd,amsmath,amssymb,twoside,showpacs,
						nofootinbib,preprintnumbers,notitlepage]
					{revtex4-1}
\usepackage[T1]{fontenc}
\usepackage{fix-cm}
\usepackage{ifthen}
\usepackage{dsfont} 
\usepackage{upgreek}
\usepackage[dvipsnames,svgnames]{xcolor}
\usepackage[citebordercolor=Apricot,
				linkbordercolor=Apricot,
				urlbordercolor=Apricot
				]{hyperref}

\newcommand{\nc}{\newcommand}

\nc{\mc}[1]{\mathcal{#1}}

\newcommand{\defeq}{:=}
\nc{\im}{\mathrm{i}}  	
\nc{\xd}{\mathrm{d}}	 	
\nc{\xD}{\mc{D}} 	

\nc{\cH}{\mc{H}} 	
\nc{\cHS}{\mc{H}^{\mathrm{S}}} 	
\nc{\cHH}{\mc{H}^{\mathrm{H}}} 	

\nc{\sphere}[1][]{\mathds{S}^{#1}}
\nc{\reals}[1][]{{\mathds{R}^{#1}}}
\nc{\complex}[1][]{{\mathds{C}^{#1}}}

\nc{\AR}{A^{\reals}}
\nc{\KD}{\mathcal{K}^\mathrm{D}}

\nc{\txg}{\mathrm{g}}
\nc{\txD}{\mathrm{D}}
\nc{\txH}{\mathrm{H}}
\nc{\txI}{\mathrm{I}}
\nc{\txR}{\mathrm{R}}
\nc{\txS}{\mathrm{S}}

\nc{\Repart}{\mathds{R}\text{e}\:}
\nc{\Impart}{\mathds{I}\text{m}\:}

\nc{\bal}[1]{\begin{align}#1\end{align}}
\nc{\bals}[1]{\begin{align*}#1\end{align*}}
\nc{\balsplit}[1]{\bal{\begin{split}#1\end{split}}}

\nc{\vc}[1]{\underline{#1}}
\nc{\lvc}[2][]{_{#1\vc #2}}

\nc{\ltx}[1]{_{\mathrm{#1}}}
\nc{\htx}[1]{^{\mathrm{#1}}}

\nc{\intval}[3]{[#1_{#2},#1_{#3}]}
\nc{\lintval}[3]{ _{ [#1_{#2},#1_{#3}] }}

\nc{\del}{\partial}
\nc{\deth}[1][]{\det\!^{#1}\,\!}

\nc{\eu}{\mathrm{e}}		
\nc{\coco}{\overline}		

\nc{\One}[1][]{{\mathds{1}_{{#1}} \,\!}}

\nc{\sign}{\,\text{sign}\,}

\nc{\unemptynumberone}[2]{\ifthenelse{\equal{#1}{}}{}{#2}}

\nc{\psistate}[3][]{\ps^{{#3}\unemptynumberone{#1}{,}#1}_{#2}\,\!}
\nc{\psS}[2][]{\psistate[#1]{#2}{\text{S}}}	
\nc{\psSar}[3][]{\psS[#1]{#2}(#3)}
\nc{\psD}[2][]{\psistate[#1]{#2}{\text{D}}}	
\nc{\psDar}[3][]{\psD[#1]{#2}(#3)}
\nc{\psH}[2][]{\psistate[#1]{#2}{\text{H}}}	
\nc{\psHar}[3][]{\psH[#1]{#2}(#3)}

\nc{\roS}[2][]{\ro^{\text{S}\unemptynumberone{#1}{,}#1}_{#2}}
\nc{\roH}[2][]{\ro^{\text{H}\unemptynumberone{#1}{,}#1}_{#2}}

\nc{\Sibar}{{\overline\Sigma}}

\nc{\Norm}[2][]{\mc N^{#1}_{#2}}

\nc{\bgl}{\left}
\nc{\biigl}{\Bigl}
\nc{\biiigl}{\biggl}
\nc{\biiiigl}{\Biggl}

\nc{\bgr}{\right}
\nc{\biigr}{\Bigr}
\nc{\biiigr}{\biggr}
\nc{\biiiigr}{\Biggr}

\nc{\bgm}{\middle}
\nc{\biigm}{\Bigm}
\nc{\biiigm}{\biggm}
\nc{\biiiigm}{\Biggm}

\nc{\bglrr}[1]{\bgl( #1 \bgr)}
\nc{\biglrr}[1]{\bigl( #1 \bigr)}
\nc{\biiglrr}[1]{\biigl( #1 \biigr)}
\nc{\biiiglrr}[1]{\biiigl( #1 \biiigr)}
\nc{\biiiiglrr}[1]{\biiiigl( #1 \biiiigr)}

\nc{\abs}[1]{\left| #1 \right|}

\nc{\fracwspace}{\hspace{0.3ex}}
\nc{\fracw}[2]{\, \frac{\fracwspace #1 \fracwspace}%
			{\fracwspace #2 \fracwspace} \,%
			}
\nc{\tfracw}[2]{\, \tfrac{\fracwspace #1 \fracwspace}%
			{\fracwspace \phantom{X^X}\hspace{-3ex} #2 \fracwspace} \,%
			}

\nc{\inpro}[3][]{\boldsymbol{\langle} #2 \boldsymbol{,} \, #3 \boldsymbol{\rangle}_{#1}}
\nc{\inproo}[3][]{\boldsymbol{\bigl\langle} #2 \boldsymbol{,} \; #3 \boldsymbol{\bigr\rangle}_{#1}}

\nc{\solinpro}[2]{\left\{#1,\,#2\right\} }

\nc{\setc}[2]{\left\{ #1 \; \middle| \; #2 \right\} }
\nc{\settc}[2]{\bigl\{ #1 \; \bigr| \; #2 \bigr\} }

\nc{\al}{{\alpha}}
\nc{\be}{{\beta}}
\nc{\ga}{{\gamma}}
\nc{\de}{{\delta}}
\nc{\ep}{{\epsilon}}
\nc{\vep}{{\varepsilon}}
\nc{\ph}{{\phi}}
\nc{\vph}{{\varphi}}
\nc{\ps}{{\psi}}
\nc{\et}{{\eta}}
\nc{\io}{{\iota}}
\nc{\ka}{{\kappa}}
\nc{\la}{{\lambda}}
\nc{\ro}{{\rho}}
\nc{\si}{{\sigma}}
\nc{\ta}{{\tau}}
\nc{\te}{{\theta}}
\nc{\vte}{{\vartheta}}
\nc{\om}{{\omega}}
\nc{\ki}{{\chi}}
\nc{\ze}{{\zeta}}

\nc{\Ga}{{\Gamma}}
\nc{\De}{{\Delta}}
\nc{\Ph}{{\Phi}}
\nc{\Ps}{{\Psi}}
\nc{\La}{{\Lambda}}
\nc{\Si}{{\Sigma}}
\nc{\Om}{{\Omega}}
\nc{\Te}{{\Theta}}
\nc{\Up}{{\Upsilon}}

\nc{\upal}{{\upalpha}}
\nc{\upvph}{{\upvarphi}}
\nc{\upte}{{\uptheta}}

\usepackage{geometry}
\begin{document}


\title{Complex structures and quantum representations\\
		for scalar QFT in curved spacetimes}
\author{Daniele Colosi\,}
\email{dcolosi@enesmorelia.unam.mx}
\author{Max Dohse\,\vspace{2mm}}
\email{max@ifm.umich.mx}
\affiliation{$^*$Escuela Nacional de Estudios Superiores, 
			Unidad Morelia,\\ 
			Universidad Nacional Aut\'onoma de M\'exico (UNAM),\\ 
			Campus Morelia, C.P.~58190, Morelia, Mexico.\vspace{1mm}
				\\
			$^\dagger$Instituto de F\'isica y Matem\'aticas (IFM-UMSNH),\\ 
			Universidad Michoacana de San Nicol\'as de Hidalgo,\\
			Ed.~C3, C.U., C.P.~58040, Morelia, Mexico.
			}


\begin{abstract}
	\noindent
	We confirm the equivalence of the Schr\"odinger representation and the holomorphic one,
	based on previous results of the General Boundary Formulation (GBF) of quantum field theory.
	On a wide class of curved spacetimes, we consider real Klein-Gordon theory in two types of regions:	
	interval regions (consisting e.g.~of a time interval times all of space), 
	and rod regions (a solid ball of space extended over all of time). 
	Using mode expansions, we provide explicit expressions for the Schr\"odinger vacuum
	(which determines this representation) and for the corresponding complex structure 
	on the space of classical solutions (which determines the holomorphic representation).
	For both representations we give the corresponding coherent states
	and calculate the generalized free transition amplitudes of the GBF, 
	which coincide and hence confirm the equivalence of the two representations.
	We also transcribe the complex structure to phase space and show that it agrees with earlier results.
\end{abstract}


\maketitle

%
\tableofcontents
%
%
\section{Introduction}
\label{sec:intro}
\noindent
In Quantum Field Theory (QFT), there exist different representations for the Hilbert space of quantum states. 
The one used most frequently is the Fock representation (using Dirac's bra-ket notation), in which a general state is a linear combination of $n$-particle states, each with defined momentum.
As for Quantum Mechanics, for QFT there is also a Schr\"odinger Representation \cite{CoCoQu:SF, CoCoQu:SF2, CoDo:Smatrix_curved, CoOe:smatrix, CoOe:letter, Hat, Jackiw, Oe:SFQ_CompObs}, in which the quantum states are represented as wave function(al)s on the space of field configurations on a 3-hypersurface (usually an equal-time hypersurface).
Alternatively, the Holomorphic Representation \cite{FadSlav:Gauge, Oe:hol, Oe:aff, Woo:geomquant} uses holomorphic wave functions on the space of classical solutions in a neighborhood of a 3-hypersurface.
The relationship between Schr\"odinger and Holomorphic Representation has been studied extensively in \cite{Oe:Sch-hol}, providing an isometric isomorphism between them.
The present article is dedicated to confirming this equivalence by giving explicit expressions for all involved quantities through mode expansions of the configurations and solutions.

We consider real, massive Klein-Gordon theory on different types of regions in a certain class of curved spacetimes, see Section \ref{sec:_Classical}.
The first type, called interval regions, is (a generalization of) the usual setting considered in QFT, namely a time interval times all of space.
A second type, called rod region, consists of a solid ball in space extended over all of time.
This type of region is very practical for situations when the usual techniques of QFT are not working, for example in AdS spacetimes \cite{Balasubramanian:1999ri, dohse:_class_AdS, DoOe:_Complex_AdS, Giddings:1999qu} and in black hole spacetimes \cite{'tHooft:1996tq} where there are no (temporal) asymptotically free states.
Since in this case the quantum states live on timelike surfaces called hypercylinders (a sphere in space extended over all of time), we need to use a formulation of Quantum Theory, which generalizes the standard formulation from equal-time hypersurfaces (respectively Cauchy surfaces) and time-interval regions to general hypersurfaces and regions.
This is precisely what the General Boundary Formulation (GBF)\cite{Oe:boundary,Oe:GBQFT,Oe:KGtl} is providing.

In this article we focus on relating the two representations, i.e.~the Schr\"odinger and the holomorphic one, 
and the GBF comes into play merely to justify states on general hypersurfaces (not necessarily Cauchy).
Therefore we only outline the principles of the GBF in a rather brief fashion.
By GBF we refer here to the version of the GBF called amplitude formalism
developed e.g.~in \cite{Oe:GBQFT,Oe:probgbf,Oe:KGtl,CoOe:smatrix},
as compared to the positive formalism of the GBF introduced in \cite{Oe:pos_GBF, Oe:loc_oper_frame}.
We emphasize first, that the GBF \emph{is not} a particular quantum theory, but an axiomatic framework on how to formulate quantum theories, inspired by Topological Quantum Field Theory. 
Roughly speaking, its axioms provide consistency conditions between the geometric (hypersurfaces and regions in spacetime) and the algebraic structures (state spaces and amplitudes) of the theory. 
In the GBF, quantum states live on hypersurfaces (of codimension one) in spacetime. 
That is, each hypersurface $\Si$ has its associated quantum state space $\cH_\Si$, which is a Hilbert space%
\footnote{By $\Sibar$ we denote the same hypersurface $\Si$ with
	opposite orientation, which also has its associated state space 
	$\cH_\Sibar$. Throughout this article, we treat both state spaces 
	as identified $\smash{\cH_\Sibar = \cH_\Si}$, 
	writing $\smash{\ps_\Sibar \defeq \coco{\ps_\Si}}$ 
	(the bar over the state denotes complex conjugation). 
	} 
in the case of Klein-Gordon theory.
In the GBF sense, the disjoint union of hypersurfaces again counts as a hypersurface, and the state space of such a hypersurface is simply the tensor product of the union's constituent hypersurfaces' state spaces.
A particular class of hypersurfaces are the boundaries $\del M$ of spacetime regions $M$ (regions have codimension zero),
and hence each region $M$'s boundary $\del M$ has its state space $\cH_{\del M}$.
As usual, we orient boundaries as pointing outwards of the enclosed regions. 
The quantum dynamics taking place inside a spacetime region $M$ is encoded by a linear amplitude map $\ro_M: \cH_{\del M} \to \complex$,
which determines an amplitude for each boundary state $\ps_{\del M} \in \cH_{\del M}$.
This amplitude map is not fixed a priori by the GBF, but depends on the particular quantum theory under consideration, which in our case happens to be real Klein-Gordon theory.

The main characteristic of the GBF consists in allowing arbitrary regions for the description of dynamics, without imposing any special form to the boundary $\del M$.
The GBF thus removes the restriction of standard QFT to Cauchy surfaces and consistently describes the dynamics of quantum fields in situations where the standard formulation is difficult to apply, while it also reproduces the results of the standard formulation of QFT.
Moreover, the GBF maintains all the constitutive properties a quantum theory should have and implements a consistent probability interpretation of its relevant structures \cite{Oe:probgbf}.
The GBF hence appears to provide a viable setting for the definition of QFT in the absence of a background metric, a fundamental desideratum for the construction of a quantum theory of gravity.

Both representations above have been applied in the GBF framework, each with a suitable quantization method.
A path integral approach \`a la Feynman naturally fits the Schr\"odinger representation, resulting in what is called Schr\"odinger-Feynman Quantization (SFQ).
For the Holomorphic Representation, a different type of functional integral is used, giving rise to the Holomorphic Quantization scheme (HQ).
The aim of this article is to contribute to the understanding of the relations between these representations and quantizations.
 
The outline of the paper is as follows. 
In Section~\ref{sec:_Classical} we specify all the relevant structures of the classical theory for the two types of spacetime regions on which the dynamics of the Klein-Gordon field will be considered, namely interval, tube and rod regions. 
For each of these regions, a mode expansion for the Klein-Gordon field is given and then used to explicitly express the main algebraic structures like the symplectic potential, the symplectic structure, the projectors on momentum and configuration subspaces, the complex structure, the real and complex inner products, and further ingredients of the quantization schemes presented in Section~\ref{sec:_quant}. 
The relation between the representations defined within the Schr\"odinger-Feynman and the Holomorphic Quantization scheme is studied in Section~\ref{sec:_Relation_SF_HOL}: 
The action of the map between the Hilbert spaces in the two representations is explicitly implemented to show the equivalence of the vacuum states, coherent states and amplitudes (for the free theory) defined in both schemes. 
Section~\ref{sec:_ComplexStructPhase} presents the action of the complex structure in phase space in terms of the quantities (modes and projectors) introduced in Section~\ref{sec:_Classical}. Finally, Section~\ref{sec:_Conclusions} briefly summarizes the results obtained.
%
%
\section{Classical Theory}
\label{sec:_Classical}
\noindent
We consider a real, massive, minimally coupled Klein-Gordon field $\ph$ 
in a 4-dimensional curved spacetime manifold%
\footnote{See \cite{CoDo:Smatrix_curved} for more details about the class of spacetimes
	which we consider and for a discussion of the well-posedness 
	when initial data is placed on timelike hypersurfaces.
	},
with Lorentzian signature and metric tensor $g_{\mu\nu}$. 
Let $y^0$ denote the time coordinate.
Then with $\si_{00} \defeq \sign g_{00}$ (making the expression independent of the metric's overall sign) the free action in a spacetime region $M$ is
\bal{		
	\label{eq:action}
	S^0_{M}(\ph) = \tfrac{1}{2} \int_M \xd^4 y\, \sqrt{|g|}\, 
	\Bigl( \si_{00}g^{\mu \nu} (\del_{y^\mu} \ph)\, 
	(\del_{y^\nu} \ph) - m^2 \ph^2 \Bigr), 
	}
where the integration is extended over the spacetime region $M$ and we use the notation $\del_{y^\mu} = \del / \del y^{\mu}$.
By $g$ we denote the determinant of the metric tensor: $g \equiv \det g_{\mu \nu}$, and $m$ indicates the mass of the field.
The action's label $0$ refers to the free theory. 
We use Einstein's sum convention in the form that a summation is understood over all Greek lowercase indices which appear exactly once as a superscript and once as a subscript in a term.
The variation of the free action yields the (homogeneous) Klein-Gordon equation as the Euler-Lagrange equation of \eqref{eq:action}:
\bal{		
	\label{eq:KG}
	\Bigl(\si_{00}\tfrac{1}{\sqrt{|g|}}\, \del_{y^\mu} \sqrt{|g|}\,
	g^{\mu \nu} \del_{y^\nu} + m^2 \Bigr)\, \ph(y)
	=0.
	}
As in \cite{CoDo:Smatrix_curved}, we suppose that the spacetime region $M$ admits a foliation 	whose leaves are hypersurfaces, described in terms of a smooth coordinate system $(\ta, \vc x)$. 
The leaves $\Si_\ta$ of the foliation are parametrized by the coordinate $\ta \in I^{(1)} \subseteq \reals$, while the coordinates on each leaf 	are denoted by $\vc x = (x^1,x^2,x^3) \in I^{(3)} \subseteq \reals[3]$.
We \emph{do not} require $\ta$ and $\vc x$ to be timelike 	and spacelike coordinates respectively, however we do require	either all leaves to be spacelike or all leaves to be timelike%
\footnote{Often $\ta$ is simply a time variable $t$ 
	or a radial variable	$r$.	
	In \cite{Oe:KGtl,Oe:hol}, also the case of $\ta$ being the
	spatial	cartesian coordinate $x^1$ has been studied.
	}.
We also assume that the metric in the coordinates $(\ta, \vc x)$  is block-diagonal:	$g^{\ta x^i} = 0 = g_{x^i \ta}$ for all $i \in \{ 1,2,3 \}$.
These assumptions certainly appear rather restrictive, but it turns out that they encompass most curved spacetimes on which QFTs are studied, including rotating, charged black hole metrics (Kerr-Newman type, in the region outside of the horizon).

In this section we introduce the spacetime regions we shall work with, 
and define and evaluate several key structures on spaces of classical solutions.
Having these expressions at hand then makes rather short work of the calculations
in the Holomorphic Quantization scheme in Section \ref{sec:_Relation_SF_HOL}.
We denote classical Klein-Gordon solutions on spacetime by the ''tall'' letters $\ph,\xi,\ze,\la(\ta,\vc x)$, whereas configurations on hypersurfaces of constant $\ta$ are denoted by the ''short'' letters $\vph, \ki, \et(\vc x)$.
%
%
\subsection{Interval regions}
\label{sec:_Classical_intervals}
\noindent
Interval regions $M\lintval\ta12 = [\ta_1,\ta_2] \times I^{(3)}$
are foliated by the leaves $\Si_\ta$ represented by $I^{(3)}\!\subseteq \reals[3]$ along the interval $[\ta_1,\ta_2]\subseteq I^{(1)}$ of the foliation parameter $\ta$.
(The standard example is a time-interval region $M\lintval t12 = [t_1,t_2] \times \reals[3]$, e.g.~in Minkowski spacetime.) 
We use the label $[\ta_1,\ta_2]$ for all quantities calculated for interval regions.
The canonical orientation of our leaves is in negative $\ta$-direction, to which we refer as backwards orientation.

Each interval region is bounded by two disjoint constant-$\ta$ hypersurfaces: $\Sigma_1$ at $\ta_1$ and $\Sigma_2$ at $\ta_2$. 
Since we orient boundaries as pointing outwards of the enclosed region,
the interval region's boundary can be written as the disjoint union
$\del M\lintval\ta12 = \Si_1\cup\Sibar_2$,
wherein the bar denotes orientation reversal.
In the case that $\ta$ is a time coordinate, and if the boundary consists of Cauchy surfaces, then the interval region is the usual setting for QFT in curved spacetime.
Independently of this, we suppose that our foliation is such that the whole spacetime can be covered by an interval region via sufficiently decreasing $\ta_1$ and increasing $\ta_2$.

Next, we introduce mode decompositions for the Klein-Gordon solutions and for the boundary field configurations, see \cite{CoDo:Smatrix_curved} for their properties.
With $\vc k $ denoting the set of three parameters (momenta) labeling the modes, we assume that there is a set of complex modes $\{U_{\vc k }(\vc x )\}$, which has the reflection property $U_{-\vc k}=\coco{U_{\vc k}}$ and forms a complete orthonormal basis in the space of field configurations on the hypersurfaces $\Si_\ta$, and also in momentum space, namely:
\bal{
	\label{eq:modenormbasis}
	\int\!\! \xd^3 k\; w_{\vc k}(\vc x)\,
	U_{\vc k }(\vc x ) \,	\coco{U_{\vc k }(\vc x ')}
	& = \delta^{(3)}(\vc x  -\vc x '),
		\\
	\label{eq:modenormbasis_k}
	\int_{\Si_\ta}\!\!\!\! \xd^3 x  \; 
	\sqrt{|g^{(3)} g^{\ta \ta}|_{\ta}}\;
	U_{\vc k }(\vc x ) \, \coco{U_{\vc k '}(\vc x )}
	& = \tilde w_{\vc k}(\ta)\,\delta^{(3)}(\vc k  -\vc k ').	
	}
Therein, $g^{(3)}$ is the determinant of the induced 3-metric on the hypersurface $\Si_\ta$.
By $\om_{\vc k}(\vc x) >0$ we denote the eigenvalue/eigenfunction of the operator $\om(\vc x)$ upon action on the basis: $w(\vc x)U_{\vc k}(\vc x) = w_{\vc k}(\vc x) U_{\vc k}(\vc x)$, ditto for $\tilde w(\ta)$ with eigenvalues $\tilde w_{\vc k}(\ta) >0$.
We require that the product of these two operators, that is of each product $w_{\vc k}(\vc x)\;\tilde w_{\vc k}(\ta)$, is positive, $\vc k$-independent, and yields the metric root $\smash{\sqrt{|g^{(3)} g^{\ta \ta}|_{\ta}}}$:
\bal{
	\label{metric_separation}
	w(\vc x)\;\tilde w(\ta)
	= \sqrt{|g^{(3)} g^{\ta \ta}|_{\ta}}.
	}
Using these modes, a field configuration $\vph(\vc x )$ on $\Si_\ta$ has the following decomposition:
\bal{
	\label{eq:modeinterval}
	\vph(\vc x ) & = \int\!\! \xd^3 k  \; 
	\vph_{\vc k } \, U_{\vc k }(\vc x ),
		&
	\vph_{\vc k} & = \int\!\! \xd^3 x\; w_{\vc k}(\vc x)\,
	\vph(\vc x)\, \coco{U_{\vc k}(\vc x)}\;.
	}
We assume that any solution of the Klein-Gordon equation can be written as (the superscripts a,b are mere labels, not indices):
\bal{		
	\label{eq:clsol}
	\phi(\ta, \vc x ) 
	= \biglrr{ X\htx a(\ta) Y\htx a}(\vc x ) 
	+ \biglrr{ X\htx b(\ta) Y\htx b}(\vc x ).
}
Therein, $X\htx a(\ta)$ and $X\htx b(\ta)$ are commuting, linear operators from the space of real-valued 
data $Y\htx a, Y\htx b$ to solutions on hypersurfaces at fixed values of $\ta$.\footnote{The expansion \ref{eq:clsol} corresponds to the assumption that the Klein-Gordon equation can be solved by separating the variables: In that case, $X_k\htx{a,b}$ represent the solution to the $\tau$-part of the Klein-Gordon equation.} 
In particular each $X\htx{a,b}(\ta)$ act as operators on a mode decomposition of $Y\htx{a,b}(\vc x)$ respectively as in \eqref{eq:modeinterval}, and we use the notation $X\htx{a,b}_{\vc k}(\ta)$ for the real eigenvalues of the corresponding operator when acting on a mode of momentum $\vc k$:
\bal{
	\label{X_action_eigenvalues_642}
	X\htx a(\ta)\,U_{\vc k}(\vc x) 
	& = X\htx a_{\vc k}(\ta)\,U_{\vc k}(\vc x),
	&
	X\htx b(\ta)\,U_{\vc k}(\vc x) 
	& = X\htx b_{\vc k}(\ta)\,U_{\vc k}(\vc x).
	}
In Section II.A of \cite{CoDo:Smatrix_curved} we show the reflection properties
\bal{
	\label{eq:Xab_k_reflect}
	X\htx{a}_{-\vc k}(\ta) & = X\htx{a}_{\vc k}(\ta),
		&
	X\htx{b}_{-\vc k}(\ta) & = X\htx{b}_{\vc k}(\ta).
	}
By linearity of the Klein-Gordon operator, we can view the functions $\left( X\htx a(\ta) Y\htx a\right)(\vc x )$ and $\left( X\htx b(\ta) Y\htx b\right)(\vc x )$ as two independent solutions of the Klein-Gordon equation \eqref{eq:KG}.
Klein-Gordon solutions on an interval region thus write as the following expansion (which we call "real" expansion, since $X\htx{a,b}_{\vc k}(\ta)$ are real):
\bal{
	\label{eq:modeinterval_sol}
	\phi(\ta,\vc x ) 
	& = \int\!\! \xd^3 k  \, \biiglrr{
		\ph\htx a_{\vc k} \,X\htx a_{\vc k}(\ta)\, U_{\vc {k}}(\vc {x})
		+ \ph\htx b_{\vc k} \,X\htx b_{\vc k}(\ta)\, U_{\vc {k}}(\vc {x})
		}.
	}
The coefficients $(\ph\htx a_{\vc k},\ph\htx b_{\vc k})$ of the real expansion determine the solution $\phi(\ta,\vc x)$ and can be recovered directly from initial data $\biglrr{\ph(T,\vc x),(\del_\ta\ph)(T,\vc x) }$ on a hypersurface $\Si_{\ta=T}$, see Section II.A in \cite{CoDo:Smatrix_curved}.
In calculations, we often need the following Wronskians, which never vanish (making them invertible) due to the linear independence of $X\htx a$ and $X\htx b$:
\bal{
	\label{eq:_Wronski_gen}
	\mc W(\ta)
	& \defeq \biglrr{X\htx a\,\del_\ta X\htx b
				-X\htx b\, \del_\ta X\htx a
				}(\ta),
		&
	\mc W_{\vc k}(\ta)
	& \defeq \biglrr{X\htx a_{\vc k}\,\del_\ta X\htx b_{\vc k}
		-X\htx b_{\vc k}\del_\ta X\htx a_{\vc k}}(\ta)\;.
	}
%
%
\subsection{Hypercylinder regions: rods and tubes}
\label{sec:_Classical_HypCyl}
\noindent
In order to define two more types of regions, we again introduce a foliation of the spacetime, defined by a smooth coordinate system $(t,r,\upte, \upvph)$. Therein, $t \in I^{(t)} \subseteq \reals$ is now a time variable and $r \in I^{(r)} \subseteq \reals$ is a radial coordinate. 
$\upte \in [ 0, \pi ]$ and $\upvph \in [0, 2 \pi)$ are angular coordinates, for which we use the collective notation $\Omega\defeq(\upte,\upvph)$ and $\xd\Om = \xd\upte\, \xd\upvph$. 
Our new regions are defined in terms of hypercylinders $\Si_r$, which are the hypersurfaces of constant radius $r$, that is: $\Si_r = I^{(t)} \times \sphere[2]_r$, wherein $I^{(t)}\subseteq\reals$ represents all of time.
Just as the surfaces of constant $\ta$ before, the hypercylinders $\Si_r$ are canonically oriented in direction of negative $r$, that is, inwards.
Here, we require the metric to be block diagonal with respect to the radial coordinate, that is: $0=g_{tr}=g_{r\upte}=g_{r\upvph}$.%
\footnote{Note that this is fulfilled e.g.~by all black hole metrics,
			including Kerr-Newman.}

As a first type, we can define tube regions: they are bounded by two concentric hypercylinders of different radii $R_1$ and $R_2$.
Hence they are a radial analogue of the interval regions defined above. 
The quantities associated to tube regions $M_{[R_1,R_2]} \defeq I^{(t)} \times [R_1,R_2] \times \sphere[2]$ are labeled by $[R_1,R_2]$
and the boundary can be written as $\del M\lintval R12 = \Si_{R_1} \cup \Sibar_{R_2}$.

The second type of region $M_R = I^{(t)} \times [0,R] \times \sphere[2]$ is called rod region and is bounded by \emph{only one} hypercylinder: $\del M_R=\Sibar_R$. 
We will use the label $R$ for the quantities associated to the rod region $M_R$. 
We assume that we can cover the whole spacetime with a rod region by sufficiently increasing $R$.

Again we introduce mode decompositions for the Klein-Gordon solutions and for the boundary field configurations. 
Since here the foliation involves the sphere $\sphere[2]$, the corresponding momenta are now discrete, and we shall denote them  simply by $l$ and $m_l$ (the subscript $l$ distinguishes $m_l$ from the field mass $m$).
Since $t$ usually takes values on the whole real line, we assume the corresponding momentum $\om$ to be continuous.
In analogy to \eqref{eq:modenormbasis}, we assume a set of complex modes $\{U_{\om l m_l\!}(t,\Om)\}$, which forms a complete orthonormal basis in the space of field configurations on the hypercylinders $\Si_r$, and also in momentum space:
\bal{\label{eq:modenormbasis_cyl}
	\int\!\! \xd\om \sum_{l,m_l} w_{\om l m_l\!}(t,\Om)\,
	U_{\om l m_l\!}(t,\Om) \,
	\coco{U_{\om l m_l\!}(t',\Om')}
	& = \delta(t\!-\! t')\,\delta^{(2)}(\Om,\Om'),
		\\
	\label{eq:modenormbasis_cyl_omlml}
	\int_{\Si_R}\!\!\!\! \xd t\,\xd\Om \sqrt{|g^{(3)} g^{rr}|_{R}}\;
	U_{\om l m_l\!}(t,\Om) \,
	\coco{U_{\om' l' m'_l\!}(t,\Om)}
	& = \tilde w_{\om l m_l\!}(R)\,\delta(\om\!-\!\om')\,\de_{ll'}\,\de_{m_l m'_l}.	
	}
Again we require the product $w\tilde w$ to yield the metric root:
\bal{		
	\label{metric_separation_r}
	w(t,\Om)\, \tilde w(R)
	= \sqrt{|g^{(3)} g^{rr}|_{R}}.
	}
Using these modes, a field configuration $\vph(t,\Om)$ on $\Si_r$ has the following decomposition:
\bal{		
	\label{eq:modeshyp}
	\vph(t,\Om) & = \int\!\!\xd\om\sum_{l,m_l}
	\vph_{\om l m_l\!}\, U_{\om l m_l\!}(t,\Om), \quad
		&
	\vph_{\om l m_l\!} 
	& = \int\!\!\xd t\,\xd\Om\; w_{\om l m_l\!}(t,\Om)\;
	\vph(t,\Om)\; \coco{U_{\om l m_l\!}(t,\Om)}\;.
	}
As for interval regions, we assume that any solution of the Klein-Gordon equation on a tube region can be written as
\bal{
	\phi(t,r,\Om) = 
	\biglrr{ X\htx a(r) Y\htx a}(t,\Om)
	+ \biglrr{ X\htx b(r) Y\htx b}(t,\Om),
	\label{eq:clsol_tube}
	}
whereas a Klein-Gordon solution on a rod region can be written as
\bal{
	\phi(t,r,\Om) = \biglrr{ X\htx a(r) Y\htx a}(t,\Om).
	\label{eq:clsol_rod}
	}
As discussed in \cite{CoDo:Smatrix_curved}, we are assuming that $X\htx a$ represents the regular solution to the radial part of the Klein-Gordon equation (e.g.~spherical Bessel functions in Minkowski spacetime), while $X\htx b$ represents the diverging solution (e.g.~spherical Neumann functions).
Klein-Gordon solutions on a tube region can then be written as a "real" expansion like \eqref{eq:modeinterval_sol}:
\bal{\label{eq:modetub}
	\phi(t,r,\Om) 
	& = \int\!\!\xd\om\sum_{l,m_l}
	\biglrr{\ph\htx a_{\om l m_l\!} \,X\htx a_{\om l m_l\!}(r)\, 
				U_{\om l m_l\!}(t,\Om)
			+ \ph\htx b_{\om l m_l\!} \,X\htx b_{\om l m_l\!}(r)\, 
			U_{\om l m_l\!}(t,\Om)
			}.
	}
The solution`s coefficients $(\ph\htx a_{\om l m_l\!},\ph\htx b_{\om l m_l\!})$  can be recovered from initial data $\biglrr{\ph(t,R,\Om),(\del_r\ph)(t,R,\Om) }$ on a hypercylinder $\Si_{R}$, see again \cite{CoDo:Smatrix_curved}.
Klein-Gordon solutions on a rod can also be expanded as a "real" expansion:
\bal{\label{eq:moderod}
	\phi(t,r,\Om) 
	& = \int\!\!\xd\om\sum_{l,m_l}
	\ph\htx a_{\om l m_l\!} \,X\htx a_{\om l m_l\!}(r)\, U_{\om l m_l\!}(t,\Om),
	}
and the momentum representation $\ph\htx a_{\om l m_l\!}$ of the solution can be recovered from Dirichlet boundary data $\ph(t,R,\Om)$ on a hypercylinder $\Si_{R}$.
%
%
\subsection{Frequency expansion}
\label{sec:_Classical_FrequencyRep}
\noindent
In this and the following sections, we write $(\ta,\vc x)$ for the coordinates and $\vc k$ for the momenta as in Section \ref{sec:_Classical_intervals} for the interval regions.
Nevertheless, everything applies as well to (tube and rod) hypercylinder regions and hypercylinder hypersurfaces $\Si_r$ (from Section \ref{sec:_Classical_HypCyl} it is clear how to adapt the notation for that case).
Starting from the "real" expansion \eqref{eq:modeinterval_sol}, we can define different expansions of the solution $\ph(\ta,\vc x)$ by using the complex linear combination
\bal{		
	\label{eq:_def_Up_k}
	\Up_{\!\vc k} (\ta) 
	:= c\htx a_{\vc k} X^a_{\vc k}(\ta) + c\htx b_{\vc k} X^b_{\vc k}(\ta),
	}
defined by choosing the two complex functions $c\htx a_{\vc k}, c\htx b_{\vc k}$ on momentum space, which in turn determine the operators $c\htx{a,b}$ through being their eigenfunctions: $c\htx a U_{\vc k}(\vc x) = c\htx a_{\vc k} U_{\vc k}(\vc x)$.
Of course, we also have $\Up(\ta) U_{\vc k}(\vc x) = \Up_{\vc k}(\ta) U_{\vc k}(\vc x)$.
In Section \ref{sec:_Relation_SF_HOL_Vacuum} we identify $c\htx{a,b}_{\vc k}$ with the functions which determine the vacuum state in the Schr\"odinger representation.
Anticipating this, we already impose the reflection properties $c\htx a\lvc k = c\htx a_{\!-\! \vc k}$ and $c\htx b\lvc k = c\htx b_{\!-\! \vc k}$, which together with \eqref{eq:Xab_k_reflect} induce the reflection property $\Up_{-\vc k}(\ta) = \Up_{\!\vc k}(\ta)$.
With this, we can write the "frequency" expansion of the solution as
\bal{		
	\label{eq:_FreqExp}
	\ph(\ta,\vc x) 
	= \int\!\! \xd^3 k\; \biiglrr{
					\ph^+_{\vc k}\, \Up_{\!\vc k} (\ta)\, U_{\vc k}(\vc x)
					+\coco{\ph^-_{\vc k}}\: \coco{\Up_{\!\vc k} (\ta)}\:
						\coco{U_{\vc k}(\vc x)}\,
					}.
	}
We call the modes $\Up_{\!\vc k} (\ta)\, U_{\vc k}(\vc x)$ positive frequency and $\coco{\Up_{\!\vc k} (\ta)}\,U_{\vc k}(\vc x)$ negative frequency, see Section V.A in \cite{CoDo:Smatrix_curved}.
For rod and tube regions, positive (negative) frequency means ingoing  (outgoing) modes.
The solution $\ph$ becomes real iff $\ph^+ = \ph^-$.
Real and frequency coefficients are related by
\renewcommand\arraystretch{1.7}
\bal{		
	\label{eq:_rel_real_freq_momreps}
	\begin{pmatrix}
		\ph\htx a_{\vc k} \\ \ph\htx b_{\vc k}
	\end{pmatrix}
	& =
	\begin{pmatrix}
		c\htx a_{\vc k}\, & \coco{c\htx a_{\vc k}}\;
			\\
		c\htx b_{\vc k}\, & \coco{c\htx b_{\vc k}}\;
	\end{pmatrix}
	\begin{pmatrix}
		\ph^+_{\vc k} \\ \coco{\ph^-_{-\vc k}}
	\end{pmatrix},\qquad
		& 
	\begin{pmatrix}
		\ph^+_{\vc k} \\ \coco{\ph^-_{-\vc k}}
	\end{pmatrix}
	& =
	\frac{-1}{2\im\, \Impart(\coco{c\htx a_{\vc k}} c\htx b_{\vc k})}
	\begin{pmatrix}
		\coco{c\htx b_{\vc k}} & -\coco{c\htx a_{\vc k}}\;
			\\
		-c\htx b_{\vc k} & \;c\htx a_{\vc k}
	\end{pmatrix}
	\begin{pmatrix}
		\ph\htx a_{\vc k} \\ \ph\htx b_{\vc k}
	\end{pmatrix}
	}
	\renewcommand\arraystretch{1}
To make these relations well defined, we need 
$\Impart(\coco{c\htx a_{\vc k}} c\htx b_{\vc k})\, \neq 0$.
In Section IV.B of \cite{CoDo:Smatrix_curved} the same requirement arises from the positivity condition for the vacuum state.
The frequency coefficients $(\ph^+_{\vc k},\ph^-_{\vc k})$ of a solution $\ph(\ta,\vc x)$ can also be obtained from initial data $\biglrr{\ph(T,\vc x),(\del_\ta\ph)(T,\vc x) }$ on a hypersurface $\Si_{\ta=T}$, again requiring $\Impart(\coco{c\htx a_{\vc k}} c\htx b_{\vc k})\, \neq 0$:
\renewcommand\arraystretch{1.6}
\bal{
	\label{recover_freqrep_interval}
	\begin{pmatrix}
		\ph^+\lvc k \\ \coco{\ph^-_{-\vc k}}
	\end{pmatrix}
	& = \int_{\Si_T}\!\! \xd^3 x \;
		w_{\vc k}(\vc x)\, \coco{U\lvc k (\vc x)}\;
		\fracw1{2\im\,\Impart(\coco{c\htx a} c\htx b)\, \mc W(T)}
		\begin{pmatrix}
		-\coco{(\del_\ta\! \Up)(T)} & \;\;\coco{\Up(T)}
			\\
		\;\;(\del_\ta\! \Up)(T) & -\Up(T)
		\end{pmatrix}
		\begin{pmatrix}
		\ph(T,\vc x) \\ (\del_\ta\ph)(T,\vc x)
		\end{pmatrix}.
	}
\renewcommand\arraystretch{1.0}%
The frequency expansion \eqref{eq:_FreqExp} is the equivalent of the Fourier transformed of a Klein-Gordon solution on Minkowski \emph{spacetime}. 
The equivalent to Fourier transformation of a configuration on a flat \emph{3-surface} is given by \eqref{eq:modeinterval}.
Taking as such configuration a solution $\ph(\ta,\vc x)$ at some fixed $\ta=T$, this can be written as
\bal{
	\label{eq:_phi_T_Fourier}
	\vph^T(\vc x) := \ph(T, \vc x) 
	& = \int\!\! \xd^3 k\; 
		\underbrace{\biiglrr{\ph^+_{\vc k}\; \Up_{\!\vc k}(T) 
										+\coco{\ph^-_{-\vc k}}\;
											\coco{\Up_{\!\vc k}(T)}
										}
						}_{\vph^{T}_{\vc k}}
		U_{\vc k}(\vc x),
	}
with the inverse given by \eqref{eq:modeinterval}:
\bal{\vph^{T}_{\vc k}
	& = \int_{\Si_T} \xd^3 x\; w(\vc x)\;
			\vph^{T}(\vc x)\, \coco{U_{\vc k}(\vc x)}.
	}
For real fields we have $\vph^T_{-\vc k} = \coco{\vph^T_{\vc k}}$.
These two relations imply that $(\vph^{T}_{\vc k} \equiv 0)$
if and only if $(\vph^{T}(\vc x) \equiv 0)$.
%
%
\subsection{Structures on spaces of classical solutions}
\label{sec:_Classical_Structures}
\noindent
In order to get a good overview of all the structures on spaces of classical solutions which later are needed for Holomorphic Quantization, we outline here their definitions and relations.
In Sections \ref{sec:_Classical_SymplecPotStruct} - \ref{sec:_Classical_R+C-InProd} we then evaluate these structures for the real and the frequency expansion.
As in Section \ref{sec:_Classical_FrequencyRep}, we use the notation of interval regions, which can easily be adapted to hypercylinders.
For each hypersurface $\Si_\ta$ of the foliation, we denote by $L_{\Si_\ta}$ the real vector space of those solutions of the Klein-Gordon equation \eqref{eq:KG} that are well defined and bounded in a neighborhood of $\Si_\ta$.
Since a region $M$'s boundary $\del M$ is a hypersurface in the GBF sense, it has its space $L_{\del M}$ of solutions that are well defined and bounded near this boundary (but not necessarily on the whole region $M$).
Using the notation and conventions of \cite{Oe:hol, Oe:Sch-hol}, we write the symplectic potential as $\te_{\Si_\ta,\xi}(\ze) =: [\xi,\ze]_{\Si_\ta}$.
As given in Eq.~(2) of \cite{Oe:aff,Oe:Sch-hol}, it is directly induced by the action, which here is \eqref{eq:action}.
The corresponding symplectic structure on $L_{\Si_\ta}$ (assumed to be  nondegenerate) according to Eq.~(19) in \cite{Oe:Sch-hol} can be written as
\bal{		
	\label{eq:_sympl_struct_from_sympl_pot}
	\om_{\Si_\ta}(\xi, \ze)
	= \tfrac{1}{2}[\xi,\ze]_{\Si_\ta} 
	- \tfrac{1}{2}[\ze,\xi]_{\Si_\ta}
	\qquad \forall\; \xi, \ze \in L_{\Si_\ta}.
	}
The symplectic potential and structure are completely determined by the classical action.
In addition to them, in the following we will also need a complex structure $J_{\Si_\ta}: L_{\Si_\ta} \to L_{\Si_\ta}$, which is linear, compatible with the symplectic structure $\om_{\Si_\ta}(J_{\Si_\ta}\,\cdot,\, J_{\Si_\ta}\,\cdot) = \om_{\Si_\ta}(\cdot,\, \cdot)$, and fulfills $J^2_{\Si_\ta} = -\One$, wherein $\One$ is the identity operator on $L_{\Si_\ta}$. 
The complex structure is not fixed by the classical theory and hence can be seen rather as a quantum object.
It induces a real, symmetric bilinear form $\txg_{\Si_\ta}$ on $L_{\Si_\ta}$ via%
\footnote{Since both are ususally denoted by the same letter, 
	we denote the spacetime metric by $g$ in italics
	and the real inner product on $L_{\Si_\ta}$ by an upright g.
	}
\bal{		
	\label{eq:_def_g}
	\txg_{\Si_\ta}(\xi, \ze)
	:= 2 \om_{\Si_\ta}(\xi, J_{\Si_\ta} \ze) 
	\qquad \forall\; \xi, \ze \in L_{\Si_\ta}.
	}
Here we require this form to be positive definite%
\footnote{As shown in \cite{Oe:fermi}, a consistent quantization
	can also be constructed with an indefinite real g-product.
	Hilbert spaces then are generalized by Krein spaces.
	}
(the complex structure is then also called positive). 
The space $L_{\Si_\ta}$ can now be completed to a real Hilbert space (which we denote by the same symbol) with the inner product $\txg_{\Si_\ta}$. 
The positive-definite, Hermitian form
\bal{		
	\label{eq:complex-inner-prod}
	\solinpro\xi\ze _{\Si_\ta}
	:= \txg_{\Si_\ta}(\xi, \ze) 
	+ 2 \im\,   \om_{\Si_\ta}(\xi,\ze) 
	\qquad \forall\; \xi, \ze \in L_{\Si_\ta} 
	}
defines a nondegenerate, sesquilinear inner product on $\smash{L_{\Si_\ta}}$, making it into complex Hilbert space. 
The momentum and configuration subspaces $M_{\Si_\ta}, N_{\Si_\ta} \subset L_{\Si_\ta}$ are defined in Eq.~(20) of \cite{Oe:Sch-hol} as
\bal{\label{eq:_MomSubspaceDef}
	M_{\Si_\ta} & \defeq \settc{\ze \in L_{\Si_\ta}}
	{[\ph,\ze]_{\Si_\ta} = 0\; \forall\; \ph \in L_{\Si_\ta}},
		\\
	\label{eq:_ConfSubspaceDef}
	N_{\Si_\ta} & \defeq \settc{\ph \in L_{\Si_\ta}}
	{[\ph,\ze]_{\Si_\ta} = 0\; \forall\; \ze \in L_{\Si_\ta}}.
	}
$M_{\Si_\ta}$ is spanned by the momentum directions in $L_{\Si_\ta}$ when viewed as its own tangent space, that is, the directions spanned by the derivative of the field in $\ta$-direction.
$N_{\Si_\ta}$ is spanned by the configuration directions in $L_{\Si_\ta}$, that is, those spanned by the field configurations on $\Si_\ta$.
By definition, $M_{\Si_\ta}$ and $N_{\Si_\ta}$  are isotropic subspaces, and we assume that they are even Lagrangian.
That is, we have the (not necessarily orthogonal) decomposition $L_{\Si_\ta} = M_{\Si_\ta}\! \oplus N_{\Si_\ta}$.

Further, $M_{\Si_\ta}\! \oplus J_{\Si_\ta}\!M_{\Si_\ta}$ is an orthogonal decomposition of $L_{\Si_\ta}$ as a real Hilbert space, wherein $J_{\Si_\ta}\!M_{\Si_\ta}$ is also Lagrangian.
The quotient space $Q_{\Si_\ta} \defeq L_{\Si_\ta}/M_{\Si_\ta}$ can be identified with the space $C_{\Si_\ta}$ of field configurations on the hypersurface $\Si_\ta$, and the quotient map is denoted by $q_{\Si_\ta}: L_{\Si_\ta} \to Q_{\Si_\ta}$.
Another crucial ingredient for the correspondence between Schr\"odinger-Feynman and Holomorphic Quantization is the unique linear map $j_{\Si_\ta}: Q_{\Si_\ta} \to (J_{\Si_\ta}\!M_{\Si_\ta}) \subset L_{\Si_\ta}$ such that $q_{\Si_\ta}\!\! \circ\! j_{\Si_\ta} = \One_Q$.

By $L_M$ we denote the real vector space of Klein-Gordon solutions that are well defined and bounded on the whole spacetime region $M$.
In the space $L_{\del M}$ of solutions near its boundary, we denote by $L\htx{int}_{\del M}$ the subspace of solutions that are not only well defined and bounded near the boundary but on the whole interior of $M$.
We consider $L_M$ and $L\htx{int}_{\del M}$ as identified here.
Remarkably, the subspace $L\htx{int}_{\del M} \subseteq L_{\del M}$ generically turns out to be Lagrangian:
The symplectic form $\omega_{\del M}$ vanishes on $L\htx{int}_{\del M}$, and it is a maximal subspace with this property.
This fact has the important consequence that $L_{\del M}$ decomposes as a direct sum $L_{\del M}=L\htx{int}_{\del M}\oplus J_{\del M} L\htx{int}_{\del M}$ over $\reals$ (for any fixed $J_{\del M}$), see Lemma 4.1 in \cite{Oe:hol}. 
Any $\la_{\del M}\in L_{\del M}$ can thus be written as the decomposition (which is unique for any fixed $J_{\del M}$):
\bal{
	\label{eq:_Decomp_R+JI}
	\la_{\del M}=\la\htx{R}+J_{\del M}\la\htx{I}
	}
wherein both $\la\htx{R},\la\htx{I}\in L\htx{int}_{\del M}$. 
(For examples, see Section~\ref{sec:_Classical_Complex_ClassAsympt_Intervals}.)
We also recall from Eq.~(81) in \cite{Oe:fermi}, that each such boundary solution $\la_{\del M}\in L_{\del M}$ has an associated element in the complexified subspace of interior solutions $(L\htx{int}_{\del M})^{\complex} \subseteq L_{\del M}^{\complex}$, which is called classical asymptotic field in \cite{ItzZub:QFT}:
\bal{
	\label{eq:_ClassAsymptoticField}
	\hat\la_M \defeq \la\htx R -\im\la\htx I.
	}
%
%
\subsection{Symplectic potential and structure}
\label{sec:_Classical_SymplecPotStruct}
\noindent
The symplectic potential $\te_{\Si_\ta,\xi}(\ph) =: [\xi,\ph]_{\Si_\ta}$ for solutions $\xi,\ph$ expanded as in \eqref{eq:_FreqExp} and using Equation (2) of \cite{Oe:aff,Oe:Sch-hol} evaluates to
\bal{		
	\label{eq:_sympl_pot_tau_x}
	[\xi,\ph]_{\Si_\ta}
	& = \si \int_{\Si_\ta}\!\!\! \xd^3 x
		\sqrt{|g^{(3)}g^{\ta\ta}|}\;
		\ph \, (\del_\ta \xi),
		\\
	\label{eq:_sympl_pot_tau_k}
	& = \si\!	\int\!\! \xd^3 k\; \tilde w_{\vc k}(\ta)
		\biiglrr{\ph^+_{\vc k}\, \xi^+_{-\vc k}\,
					\Up_{\!\vc k} (\ta)\, \del_\ta\Up_{\!\vc k} (\ta)
					+ \ph^+_{\vc k}\, \coco{\xi^-_{\vc k}}\,
					\Up_{\!\vc k} (\ta)\, \coco{\del_\ta\Up_{\!\vc k} (\ta)}
		\\
	& \hspace{23mm}
					+\coco{\ph^-_{\vc k}}\: \coco{\xi^-_{-\vc k}}\:
					\coco{\Up_{\!\vc k} (\ta)}\:
					\coco{\del_\ta\Up_{\!\vc k} (\ta)}
					+\coco{\ph^-_{\vc k}}\, \xi^+_{\vc k}\,
					\coco{\Up_{\!\vc k} (\ta)}\, \del_\ta\Up_{\!\vc k} (\ta)
					}.
		\notag
	}
Therein, $\si := \sign g_{00}\, \sign g^{\ta\ta}$, and $\Si_\ta$ is oriented in negative $\ta$-direction (backwards).
With \eqref{eq:_sympl_struct_from_sympl_pot}, the symplectic structure then takes its standard form, compare for example to Eq.~(7.2.4) in \cite{Woo:geomquant}:
\bal{		
	\label{eq:_sympl_struct_tau_x}	
	\om_{\Si_\ta} \biglrr{\xi,\ze}
	& = -\tfrac\si2 \int_{\Si_\ta}\!\!\! \xd^3 x
		\sqrt{|g^{(3)}g^{\ta\ta}|}\;
		\biglrr{\xi \del_\ta \ze - \ze \del_\ta \xi},
		\\
	\label{eq:_sympl_struct_tau_k_+-}
	& = \im\si\! 	\int\!\! \xd^3 k\; 
		\tilde w_{\vc k}(\ta)\mc W_{\vc k} (\ta)\,
		\Impart(\coco{c\htx a_{\vc k}} c\htx b_{\vc k})
		\biiglrr{\xi^+_{\vc k}\,\coco{\ze^-_{\vc k}}
					-\coco{\xi^-_{\vc k}}\, \ze^+_{\vc k}\,
					},
		\\
	\label{eq:_sympl_struct_tau_k_ab}
	& = -\tfrac\si2	\int\!\! \xd^3 k\; 
		\tilde w_{\vc k}(\ta)\mc W_{\vc k} (\ta)\,
		\biiglrr{\xi\htx a_{\vc k}\,\ze\htx b_{-\vc k}
					-\xi\htx b_{\vc k}\, \ze\htx a_{-\vc k}\,
					}.
	}
The symplectic potential and structure of the same hypersurface but with opposite orientation have opposite sign: $\smash{\om_{\Sibar_\ta} = -\om_{\Si_\ta}}$.
It can be shown \cite{Woo:geomquant} that the value of $\smash{\om_{\Si_\ta} (\xi,\ze)}$ is
independent of the hypersurface and thus independent of $\ta$,
implying that the weighted Wronskian $\tilde w_{\vc k}(\ta)\,\mc W_{\vc k} (\ta)$ must also be independent of $\ta$.
%
%
With the form \eqref{eq:_sympl_pot_tau_x} of the symplectic potential, the momentum subspace $M_{\Si_\ta}$ of \eqref{eq:_MomSubspaceDef} must consist of those real solutions that vanish on $\Si_\ta$:
\bal{
	\label{eq:_MomSpace_Vanish_SiT}
	M_{\Si_\ta} 
	\defeq \settc{\ph \in L_{\Si_\ta}}
						{[\xi,\ph]_{\Si_\ta} = 0\; 
						\forall\; \xi \in L_{\Si_\ta}
						}
	\,=\, \settc{\ph \in L_{\Si_\ta}}
		{\ph|_{\Si_\ta} \equiv 0}.
	}
This makes it clear that the quotient space $Q_{\Si_\ta}$ can be identified with the space of field configurations on $\Si_\ta$, that is: $Q_{\Si_\ta}\defeq L_{\Si_\ta}/M_{\Si_\ta}	\,=\, 
\setc{\vph \defeq \ph|_{\Si_\ta}}{\ph \in L_{\Si_\ta}}$.
It's quotient map $q_{\Si_\ta}: L_{\Si_\ta} \to Q_{\Si_\ta}$ thus is defined simply by $q_{\Si_\ta}\!:\; \ph \mapsto \vph \defeq \ph|_{\Si_\ta}$.
From \eqref{eq:_sympl_struct_tau_x} we read off that the subspace $M_{\Si_\ta}$ is indeed isotropic: $\om_{\Si_\ta}(\et,\ze)=0$ for all $\et,\ze \in M_{\Si_\ta}$.
Moreover, it is even coisotropic: $\om_{\Si_\ta}(\xi,\ze)=0$ for all $\xi,\ze \in M^\perp_{\Si_\ta}$, wherein $M^\perp_{\Si_\ta} := 
\setc{\ph \in L_{\Si_\ta}}{\om_{\Si_\ta}(\ph,\xi) = 0\; \forall\; \xi \in M_{\Si_\ta}}$ is the symplectic complement of $M_{\Si_\ta}$.
That is, our above assumption of $M_{\Si_\ta}$ being Lagrangian is true for Klein-Gordon theory.
From \eqref{eq:_phi_T_Fourier} we deduce that the momentum subspace $M_{\Si_T}$ is spanned by solutions $\ph$ with $\vph^{T}_{\vc k} \equiv 0$, which is equivalent to their frequency coefficients fulfilling
\bal{
	\label{eq:_ph+_ph-_UpUp}
	\ph^+_{\vc k} = - \coco{\ph^-_{-\vc k}}\,
	\tfracw{\coco{\Up_{\!\vc k}(T)}}{\Up_{\!\vc k}(T)}.
	}
Hence the solutions $\ph \in M_{\Si_T}$ have the frequency expansion
\bal{
	\ph(\ta,\vc x) & = \int\!\! \xd^3 k\; 
			\biiglrr{-\coco{\ph^-_{-\vc k}}
					\tfracw{\coco{\Up_{\!\vc k}(T)}}{\Up_{\!\vc k}(T)}
					U_{\vc k}(\vc x)\, \Up_{\!\vc k} (\ta) 
					+ \coco{\ph^-_{\vc k}}\: \coco{U_{\vc k}(\vc x)}\:
					\coco{\Up_{\!\vc k} (\ta)}
					}
		\notag
		\\
	\label{eq:_ph_MomSubspace_FreqExp}
	& = \int\!\! \xd^3 k\; \coco{\ph^-_{\vc k}}\:
		\coco{U_{\vc k}(\vc x)}\:
		\biiglrr{\coco{\Up_{\!\vc k} (\ta)} -\Up_{\!\vc k} (\ta)
					\tfracw{\coco{\Up_{\!\vc k}(T)}}{\Up_{\!\vc k}(T)} 
					}.
	}
The last line directly reveals that they vanish at $\ta = T$, and from \eqref{eq:_ph+_ph-_UpUp} it follows that the integrand with $\vc k$ replaced by $-\vc k$ is the complex-conjugate of the original one, that is: all the $\ph \in M_{\Si_T}$ are indeed real-valued.
Thus we can drop the minus-superscript: $\ph^-_{\vc k} \to \ph_{\vc k}$.
Further, with \eqref{eq:_sympl_pot_tau_x} the configuration subspace $N_{\Si_\ta}$ of \eqref{eq:_ConfSubspaceDef} must consist of those real solutions with vanishing derivative on $\Si_\ta$:
\bal{
	\label{eq:_ConfSpace_Vanish_SiT}
	N_{\Si_\ta} 
	\defeq \settc{\xi \in L_{\Si_\ta}}
						{[\xi,\ph]_{\Si_\ta} = 0\; 
						\forall\; \ph \in L_{\Si_\ta}
						}
	\,=\, \settc{\xi \in L_{\Si_\ta}}
		{\del_\ta\xi|_{\Si_\ta} \equiv 0}.
	}
As above for $M_{\Si_\ta}$, we can show that $N_{\Si_\ta}$ is indeed Lagrangian for Klein-Gordon theory.
The configuration subspace $N_{\Si_T}$ is spanned by solutions $\ph$ with 
\bal{
	\label{eq:_ph+_ph-_DotUpDotUp}
	\ph^+_{\vc k} = - \coco{\ph^-_{-\vc k}}\,
	\tfracw{\coco{\del_\ta\!\Up_{\!\vc k}(T)}}
			{\del_\ta\!\Up_{\!\vc k}(T)}.
	}
Setting $\ph(T,\vc x) \equiv 0$ in \eqref{recover_freqrep_interval}, and writing $\del_\ta\!\ph$ with the frequency expansion \eqref{eq:_FreqExp}, we find the projector $P_M$ on the subspace $M_{\Si_T}$, while setting $\del_\ta\ph(T,\vc x) \equiv 0$ yields $P_N$:
\renewcommand\arraystretch{1.5}
\bal{
	\label{eq:_PM_freq}
	\begin{pmatrix}
		(P_M \ph)^+_{\vc k} \\ \coco{(P_M \ph)^-_{-\vc k}}
	\end{pmatrix}
	& =
	\fracw{1}{2\im\, \Impart(\coco{c\htx a_{\vc k}} c\htx b_{\vc k})\,
					\mc W_{\vc k}(T)}
	\begin{pmatrix}
		\;\;\coco{\Up_{\vc k}(T)}\, \del_\ta\!\Up_{\vc k}(T) &
		\;\;\coco{\Up_{\vc k}(T)}\: \coco{\del_\ta\!\Up_{\vc k}(T)} \\
		-\Up_{\vc k}(T)\, \del_\ta\!\Up_{\vc k}(T) &
		-\Up_{\vc k}(T)\, \coco{\del_\ta\!\Up_{\vc k}(T)}
	\end{pmatrix}
	\begin{pmatrix}
		\ph^+_{\vc k} \\ \coco{\ph^-_{-\vc k}}
	\end{pmatrix},
		\\ 
	\begin{pmatrix}
		(P_N \ph)^+_{\vc k} \\ \coco{(P_N \ph)^-_{-\vc k}}
	\end{pmatrix}
	& =
	\fracw{1}{2\im\, \Impart(\coco{c\htx a_{\vc k}} c\htx b_{\vc k})\,
					\mc W_{\vc k}(T)}
	\begin{pmatrix}
		-\Up_{\vc k}(T)\, \coco{\del_\ta\!\Up_{\vc k}(T)} &
		-\coco{\Up_{\vc k}(T)}\: \coco{\del_\ta\!\Up_{\vc k}(T)} \\
		\;\;\Up_{\vc k}(T)\, \del_\ta\!\Up_{\vc k}(T) &
		\;\;\coco{\Up_{\vc k}(T)}\, \del_\ta\!\Up_{\vc k}(T)
	\end{pmatrix}
	\begin{pmatrix}
		\ph^+_{\vc k} \\ \coco{\ph^-_{-\vc k}}
	\end{pmatrix}.
	}
\renewcommand\arraystretch{1.0}%
It is easy to verify that the solution $P_M\ph$ fulfills \eqref{eq:_ph+_ph-_UpUp} and that $P_N$ satisfies \eqref{eq:_ph+_ph-_DotUpDotUp}, that is, $P_M\ph \in M_{\Si_T}$ and $P_N\ph \in N_{\Si_T}$.
We can also quickly check the idempotency: $P_M^2 = P_M$ and $P_N^2 = P_N$, which proves that these operators are indeed projectors.
Moreover $P_M + P_N = \One$, showing that $L_{\Si_T} = M_{\Si_T}\! \oplus N_{\Si_T}$ really holds true.
In order to see whether they are orthogonal projectors, which is equivalent to being self-adjoint, we need to check whether they fulfill $\solinpro{P_M\xi}{\ze}_{\Si_T}\! = \solinpro{\xi}{P_M\ze}_{\Si_T}\!$ and $\solinpro{P_N\xi}{\ze}_{\Si_T}\! = \solinpro{\xi}{P_N\ze}_{\Si_T}\!$ for all $\xi,\ze \in L_{\Si_T}$.
Using \eqref{eq:_complex_inpro_+-}, we see that this \emph{is not} the case.
This was foreseen in \cite{Oe:Sch-hol}: They are orthogonal only if $N_{\Si_T} = (J_{\Si_T}\! M_{\Si_T})$, which is not the case here,
and hence they are oblique projectors.
%
%
\subsection{Complex structure}
\label{sec:_Classical_ComplexStruct}
\noindent
The next object to consider is the complex structure $J_{\Si_\ta}$ on $L_{\Si_\ta}$. 
Just as the above, it changes sign under orientation reversal: $J_{\Sibar_\ta} = -J_{\Si_\ta}$.
Since $J_{\Si_\ta}$ maps solutions to solutions, we can write its action in the real expansion \eqref{eq:modeinterval_sol} 
and the frequency expansion \eqref{eq:_FreqExp} as follows:
\bal{
	\label{eq:modeinterval_Jab}
	\biglrr{J_{\Si_\ta}\phi}(\ta,\vc x ) 
	& = \int\!\! \xd^3 k  \, \biiglrr{
		(J_{\Si_\ta}\phi)\htx a_{\vc k}\,
		X\htx a_{\vc k}(\ta)\, U_{\vc {k}}(\vc {x})
		+ (J_{\Si_\ta}\phi)\htx b_{\vc k}\,
		X\htx b_{\vc k}(\ta)\, U_{\vc {k}}(\vc {x})
		},
		\\
	\label{eq:modeinterval_J+-}
	& = \int\!\! \xd^3 k  \, \biiglrr{
		(J_{\Si_\ta}\phi)^+_{\vc k}\,
		\Up_{\vc k}(\ta)\, U_{\vc {k}}(\vc {x})
		+ \coco{(J_{\Si_\ta}\phi)^-_{\vc k}}\:
		\coco{\Up_{\vc k}(\ta)}\: \coco{U_{\vc {k}}(\vc {x})}
		}.
	}
The action of $J_{\Si_\ta}$ on the solution is thus determined 
by its action on the coefficients of the real respectively 
frequency expansion of the solution.
For the real expansion, a complex structure arises from 
the functions $c\htx{a,b}_{\vc k}$:
\renewcommand\arraystretch{1.5}
\bal{		
	\label{eq:_J_c_ab}
	\begin{pmatrix}
		(J_{\Si_\ta}\phi)\htx a_{\vc k} 
			\\
		(J_{\Si_\ta}\phi)\htx b_{\vc k}
	\end{pmatrix}
	& = J\htx{ab}
	\begin{pmatrix}
		\phi\htx a_{\vc k} 
			\\
		\phi\htx b_{\vc k}
	\end{pmatrix},
	&
	J\htx{ab}
	&=
	\frac1{\Impart (\coco{c\htx a_{\vc k}} c\htx b_{\vc k})}
	\begin{pmatrix}
		\Repart (\coco{c\htx a_{\vc k}} c\htx b_{\vc k})
			&
		-|c\htx a_{\vc k}|^2
			\\
		|c\htx b_{\vc k}|^2
			&
		-\Repart (\coco{c\htx a_{\vc k}} c\htx b_{\vc k})
	\end{pmatrix}.
	}
	\renewcommand\arraystretch{1.0}%
We could also have chosen the negative of this instead, however it turns out later that the above choice leads to a positive definite real g-product.
It is easy to check that $J^2_{\Si_\ta} = -\One$.
Due to $c\htx{a,b}_{-\vc k} = c\htx{a,b}_{\vc k}$ and the above matrix being real-valued, we get: if $\ph\htx a_{-\vc k} = \coco{\ph\htx a_{\vc k}}$, then also $(J_{\Si_\ta}\ph)\htx a_{-\vc k} = \coco{(J_{\Si_\ta}\ph)\htx a_{\vc k}}$, and ditto for $\ph\htx b$.
That is, if $\ph$ is real, then $J_{\Si_\ta}\ph$ is also real.
The compatibility with the symplectic structure \eqref{eq:_sympl_struct_tau_k_ab} can be verified as follows.
First note, that we can write
\bals{ 
	\xi\htx a_{\vc k}\,\ze\htx b_{-\vc k}
	-\xi\htx b_{\vc k}\, \ze\htx a_{-\vc k}
	&=
	\biiglrr{\xi\htx a_{\vc k} \quad \xi\htx b_{\vc k}}
	\begin{pmatrix}
		\;\;0 & 1 \\
		-1 & 0
	\end{pmatrix}
	\begin{pmatrix}
		\ze\htx a_{-\vc k} \\ \ze\htx b_{-\vc k}
	\end{pmatrix}.
	}
When evaluating $\om_{\Si_\ta}(J_{\Si_\ta}\xi,J_{\Si_\ta}\ze)$,
we obtain just the same (with $\htx T$ denoting the transposed matrix):
\bals{
	(J_{\Si_\ta}\xi)\htx a_{\vc k}\,(J_{\Si_\ta}\ze)\htx b_{-\vc k}
	-(J_{\Si_\ta}\xi)\htx b_{\vc k}\, (J_{\Si_\ta}\ze)\htx a_{-\vc k}
	&=
	\biiglrr{(J_{\Si_\ta}\xi)\htx a_{\vc k} \quad
				(J_{\Si_\ta}\xi)\htx b_{\vc k}
				}
	\begin{pmatrix}
		\;\;0 & 1 \\
		-1 & 0
	\end{pmatrix}
	\begin{pmatrix}
		(J_{\Si_\ta}\ze)\htx a_{-\vc k} \\ 
		(J_{\Si_\ta}\ze)\htx b_{-\vc k}
	\end{pmatrix},
		\\
	& =
	\biiglrr{\xi\htx a_{\vc k} \quad \xi\htx b_{\vc k}}
	(J\htx{ab})\htx T
	\begin{pmatrix}
		\;\;0 & 1 \\
		-1 & 0
	\end{pmatrix}
	J\htx{ab}
	\begin{pmatrix}
		\ze\htx a_{-\vc k} \\ \ze\htx b_{-\vc k}
	\end{pmatrix},
		\\
	& =
	\biiglrr{\xi\htx a_{\vc k} \quad \xi\htx b_{\vc k}}
	\begin{pmatrix}
		\;\;0 & 1 \\
		-1 & 0
	\end{pmatrix}
	\begin{pmatrix}
		\ze\htx a_{-\vc k} \\ \ze\htx b_{-\vc k}
	\end{pmatrix},	
	}
which shows the compatibility condition $\om_{\Si_\ta}(J_{\Si_\ta}\xi,J_{\Si_\ta}\ze) = \om_{\Si_\ta}(\xi,\ze)$.
The relation between the matrix elements of $J\htx{ab}$ and the functions $c\htx{a,b}_{\vc k}$ has been studied in Section 3.3.7 of \cite{dohse:_PhD_thesis}. Suppressing momentum labels $\vc k$, we write $c\htx a = r\ltx a\, \eu^{\im \upal\ltx a}$ and $c\htx b = r\ltx b\, \eu^{\im \upal\ltx b}$ with $r\ltx{a,b} > 0$ and $\upal\ltx{a,b} \in \reals$.
Then, $\Impart (\coco{c\htx a_{\vc k}}c\htx b_{\vc k}) = r\ltx a r\ltx b \sin(\upal\ltx b \!-\! \upal\ltx a) \neq 0$ precisely if $\De\upal \defeq (\upal\ltx b\!-\!\upal\ltx a) \neq \pm n \pi$ for any $n \in \mathds{N}_0$.
Comparing to \eqref{eq:_J_c_ab}, with $Q_r\defeq r\ltx a / r\ltx b$ we obtain
\bal{
	\label{Jab_r_phi}
	(J\htx{ab})_{11} & =\tfracw{\cos \De\upal}{\sin\De\upal},
		&
	(J\htx{ab})_{12} & =-Q_r/\sin \De\upal,
		&
	(J\htx{ab})_{21} & =+Q_r^{-1}/\sin \De\upal.	
	}
That is, the matrix $J\htx{ab}$ is determined completely by the two real functions $Q_r$ and $\De\upal$.
The inverse relations are provided by
$Q_r =\sqrt{-(J\htx{ab})_{12}/(J\htx{ab})_{21}}$ and
$\De\upal =\arcsin \sqrt{-1/((J\htx{ab})_{12}(J\htx{ab})_{21})}$.
Therefore all choices of $c\htx{a,b}_{\vc k}$ that lead to the same $Q_r$ and the same $\De\upal$ (up to multiples of $2\pi$), also lead to the same matrix $J\htx{ab}$ and hence to the same complex structure.
Since the complex structures and vacuum states in the Schr\"odinger representation are in a one-to-one correspondence, all choices of $c\htx{a,b}_{\vc k}$ that lead to the same complex structure also lead to the same vacuum state, see Section \ref{sec:_Relation_SF_HOL_Vacuum}.
We could thus fix $c\htx a_{\vc k} \equiv 1$, such that the complex structure and the Schr\"odinger vacuum are completely determined by $c\htx b_{\vc k}$.
Alternatively, we could fix $c\htx b_{\vc k} \equiv 1$, which simplifies the formulas for rod regions, and then complex structure and the vacuum are determined by $c\htx a_{\vc k}$.
In this article, we shall not use these ''asymmetric'' choices, and rather stick to the ''symmetric'' notation of writing both $c\htx a_{\vc k}$ and $c\htx b_{\vc k}$ explicitly, because the formulas for the asymmetric choices can be read off from the symmetric ones, but not vice versa.

Using \eqref{eq:_rel_real_freq_momreps} we can calculate the action of $J_{\Si_\ta}$ also in the frequency expansion:
\renewcommand\arraystretch{1.6}
\bal{		
	\begin{pmatrix}
		(J_{\Si_\ta}\ph)^+_{\vc k} \\ 
		\coco{(J_{\Si_\ta}\ph)^-_{-\vc k}}
	\end{pmatrix}
	& =
	\frac{-1}{2\im\, \Impart(\coco{c\htx a_{\vc k}} c\htx b_{\vc k})}
	\begin{pmatrix}
		\coco{c\htx b_{\vc k}} & -\coco{c\htx a_{\vc k}}\;
			\\
		-c\htx b_{\vc k} & \;c\htx a_{\vc k}
	\end{pmatrix}
	\begin{pmatrix}
		(J_{\Si_\ta}\ph)\htx a_{\vc k} \\ 
		(J_{\Si_\ta}\ph)\htx b_{\vc k}
	\end{pmatrix}.
	}
	\renewcommand\arraystretch{1.0}%
Independently of which functions $c\htx{a,b}_{\vc k}$ we choose, with \eqref{eq:_J_c_ab} this results in the simple action
\bal{
	\label{eq:_J_+-}
	(J_{\Si_\ta} \ph)^\pm_{\vc k}
	= -\im\,\ph^\pm_{\vc k}.
	}
It is straightforward to check that this action also fulfills $J^2_{\Si_\ta} = -\One$ and is compatible with the symplectic structure \eqref{eq:_sympl_struct_tau_k_+-}.
Further, if $\ph^+_{\vc k} = \ph^-_{\vc k}$, then also $(J_{\Si_\ta}\ph)^+_{\vc k} = (J_{\Si_\ta}\ph)^-_{\vc k}$.
That is: if $\ph$ is real, then $J_{\Si_\ta}\ph$ is real, too.
Finally, $J_{\Si_\ta}$ also conserves the positive and negative frequency parts: $\ph^+_{\vc k}$ encodes the positive frequency part of $\ph$, and $-\im\ph^+_{\vc k}$ is the positive frequency part of $J_{\Si_\ta}\ph$, ditto for the negative frequency parts.
We can now write down the projectors $P^\mp := \tfrac12 (\One \pm \im J_{\Si_\ta})$ on the subspaces of positive and negative frequencies: \eqref{eq:_J_+-} induces the simple form
\renewcommand\arraystretch{1.5}
\bal{\label{eq:_P+-}
	\begin{pmatrix}
		(P^+ \ph)^+_{\vc k} \\ \coco{(P^+ \ph)^-_{-\vc k}}
	\end{pmatrix}
	& =
	\begin{pmatrix}
		0 & 0 \\ 0 & 1
	\end{pmatrix}
	\begin{pmatrix}
		\ph^+_{\vc k} \\ \coco{\ph^-_{-\vc k}}
	\end{pmatrix}
	=
	\begin{pmatrix}
		0 \\ \coco{\ph^-_{-\vc k}}
	\end{pmatrix}\!,\qquad
		& 
	\begin{pmatrix}
		(P^- \ph)^+_{\vc k} \\ \coco{(P^- \ph)^-_{-\vc k}}
	\end{pmatrix}
	& =
	\begin{pmatrix}
		1 & 0 \\ 0 & 0
	\end{pmatrix}
	\begin{pmatrix}
		\ph^+_{\vc k} \\ \coco{\ph^-_{-\vc k}}
	\end{pmatrix}
	=
	\begin{pmatrix}
		\ph^+_{\vc k} \\ 0
	\end{pmatrix}\!.
	}
\renewcommand\arraystretch{1.0}%
%
%
\subsubsection{Momentum subspace, complement and associated maps}
\label{sec:_Classical_Complex_MomSubspace}
\noindent
The orthogonal complement $J_{\Si_T}\!M_{\Si_T}$ of $M_{\Si_T}$ in $L_{\Si_T}$ is determined by the complex structure \eqref{eq:_J_+-}.
Applying it to the frequency expansion \eqref{eq:_FreqExp} and then inserting \eqref{eq:_ph+_ph-_UpUp} yields the frequency expansion for solutions $\xi \in (J_{\Si_T}\!M_{\Si_T})$:
\bal{
	\xi(\ta,\vc x) 
	& = \int\!\! \xd^3  k\; 
		\biiglrr{\im \,\coco{\xi^-_{-\vc k}}
				\tfracw{\coco{\Up_{\!\vc k}(T)}}{\Up_{\!\vc k}(T)}
				U_{\vc k}(\vc x)\; \Up_{\!\vc k} (\ta) 
				+ \im\,\coco{\xi^-_{\vc k}}\: \coco{U_{\vc k}(\vc x)}\:
				\coco{\Up_{\!\vc k} (\ta)}
				},
		\notag
		\\
	\label{eq:_J_ph_J_MomSubspace_FreqExp}
	& = \int\!\! \xd^3  k\; \im\, \coco{\xi^-_{\vc k}}\;
		\coco{U_{\vc k}(\vc x)}\;
		\biiglrr{ \coco{\Up_{\!\vc k} (\ta)} + \Up_{\!\vc k} (\ta)
					\tfracw{\coco{\Up_{\!\vc k}(T)}}{\Up_{\!\vc k}(T)}
					}.
	}
Comparing this to \eqref{eq:_ph+_ph-_DotUpDotUp}, we observe that for Klein-Gordon theory the subspaces $N_{\Si_T}$ and $J_{\Si_T}\!M_{\Si_T}$  \emph{do not} coincide, as already taken into account in Sections 3.1 and 5.3 of \cite{Oe:Sch-hol}.
The fields $\xi \in (J_{\Si_T}\!M_{\Si_T})$ are real by the same argument holding for the $\ph \in M_{\Si_T}$, and thus again we drop the minus-superscript: $\xi^-_{\vc k} \to \xi_{\vc k}$.
The configuration $\ki^T (\vc x) \defeq \xi(T,\vc x)$ of the solution $\xi(\ta,\vc x)$ then takes the form
\bal{
	\label{eq:_vph_JT_FreqExp}
	\ki^{T} (\vc x) 
	= \int\!\! \xd^3 k\;	\underbrace{2\im\,\coco{\Up_{\!\vc k}(T)}\, 
		\coco{\xi_{\vc k}}}_{\ki^{T}_{\vc k}} \coco{U_{\vc k}(\vc x)}\,.
	}
The inverse transformation \eqref{eq:modeinterval} provides the coefficients of the frequency expansion of a solution $\xi\in(J_{\Si_T} M_{\Si_T})$ as a function of its configuration $\ki^T(\vc x)$:
\bal{
	\label{eq:_FreqRep_xi_ki_JM}
	\coco{\xi_{\vc k}} 
	\defeq \int_{\Si_T}\!\!\! \xd^3 x\;
	\tfracw{w_{\vc k} (\vc x)}{2\im\,\coco{\Up_{\!\vc k}(T)}}
	U_{\vc k}(\vc x)\,\ki^{T} (\vc x) \;.
	}
With this relation we can now define the map $j_{\Si_T} : Q_{\Si_T} 
\to (J_{\Si_T}\!M_{\Si_T}) \subset L_{\Si_T}$ through
$j_{\Si_T}:\ki^T \mapsto \xi$, wherein $\xi$ is defined as in \eqref{eq:_J_ph_J_MomSubspace_FreqExp} with \eqref{eq:_FreqRep_xi_ki_JM}.
By construction we now have the image of $j_{\Si_T}$ in the subspace $ J_{\Si_T}\!M_{\Si_T}$.
Moreover, $j_{\Si_T}$ is linear, and with \eqref{eq:modenormbasis} it is easily verified that for all configurations $\ki^T(\vc x)$ on $\Si_T$ we have indeed
\bal{
	(q_{\Si_T} \!\circ\! j_{\Si_T} \ki^T) (\vc x)
	= \xi(T,\vc x)
	= \ki^T (\vc x)\;,
	}
that is: $q_{\Si_T} \!\circ\! j_{\Si_T} = \One_{Q}$.
Our map $j_{\Si_T}$ thus fulfills all three properties required for it in the end of Section 3.1 of \cite{Oe:Sch-hol}.
Using either $(J_{\Si_T})^2 = - \One$ or again \eqref{eq:_J_+-}, we readily obtain the last property for $\xi \in (J_{\Si_T}M_{\Si_T})$:
\bal{
	\label{eq:_J_xi_M}
	(J_{\Si_T} \xi) (\ta,\vc x)
	& = -\int\!\! \xd^3  k\; \coco{\xi_{\vc k}}\:
		\coco{U_{\vc k}(\vc x)}\;
		\biiglrr{\coco{\Up_{\!\vc k} (\ta)} - \Up_{\!\vc k} (\ta)
				\tfracw{\coco{\Up_{\!\vc k}(T)}}{\Up_{\!\vc k}(T)} 
				}.
	}
%
%
\subsubsection{Classical asymptotic field: Interval regions}
\label{sec:_Classical_Complex_ClassAsympt_Intervals}
\noindent
The complex structure \eqref{eq:_J_+-} is independent of $\ta$ and allows us to explicitly calculate the classical asymptotic field \eqref{eq:_ClassAsymptoticField} for interval and tube regions, where we denote it by
\bal{\hat\la_{12} : =\la\htx{R}_{12}-\im \la\htx{I}_{12}. }
For interval regions $M\lintval\ta12$, a real-valued solution $\la_{\del12}$ near its boundary $\Si_{\ta_1} \cup\, \Sibar_{\ta_2}$ consists of two real-valued independent solutions: $\ze$ near $\Si_{\ta_1}$, and $\xi$ near $\Si_{\ta_2}$.
That is: $\la_{\del12} = (\ze, \xi)$.
The complex structure on the boundary also has two components here: $J_{\del12} = (J_{\Si_{\ta_1}}, -J_{\Si_{\ta_1}})$, wherein the minus sign is due to the opposite orientation of $\Sibar_{\ta_2}$.
Then, decomposition \eqref{eq:_Decomp_R+JI} can be written as
\bal{
	\la_{\del12} 
	= \begin{pmatrix}
		\ze \\ \xi
		\end{pmatrix}
	= \begin{pmatrix} 
		\la\htx{R}_{12} \\ \la\htx{R}_{12}
		\end{pmatrix}
		+
		\begin{pmatrix}
		\;\;J_{\Si_{\ta_1}\!}\la\htx{I}_{12}\\
		 \!-J_{\Si_{\ta_1}\!}\la\htx{I}_{12}
		\end{pmatrix},
	}
which is solved by
\bal{
	\label{eq:_laRI12}
	\la\htx R_{12} 
	& = \tfrac12\,(\ze+\xi),
		&
	\la\htx I_{12} 
	& = \tfrac12\,(-J_{\Si_{\ta_1}}\!\ze + J_{\Si_{\ta_1}}\!\xi).
	}
Using the frequency expansion \eqref{eq:_FreqExp} for $\ze,\xi$ and the complex structure \eqref{eq:_J_+-}, we quickly obtain the frequency expansion of the classical asymptotic field:
\bal{		
	\label{eq:_laHat12_FreqExp}
	\hat\la_{12}(\ta,\vc x) 
	= \int\!\! \xd^3 k\; \biiglrr{
					\ze^+_{\vc k}\, \Up_{\!\vc k} (\ta)\, U_{\vc k}(\vc x)
					+\coco{\xi^-_{\vc k}}\: \coco{\Up_{\!\vc k} (\ta)}\:
						\coco{U_{\vc k}(\vc x)}\,
					}.
	}
For interval regions we see that the positive frequency part of $\hat\la_{12}$ is determined by the positive frequency part of the solution $\ze$ near $\Si_{\ta_1}$, and its negative frequency part by the negative frequency part of the solution $\xi$ near $\Sibar_{\ta_2}$.
Defining the following two complex configurations as
\bal{		
	\et(\vc x) 
	& := \int\!\! \xd^3 k\; \ze^+_{\vc k}\, 
			U_{\vc k}(\vc x),
		&
	\coco{\ki(\vc x)}
	& := \int\!\! \xd^3 k\; \coco{\xi^-_{\vc k}}\:
			\coco{U_{\vc k}(\vc x)},
	}
we find that the above $\hat\la_{12}$ coincides with the $\hat\la_{12}$
from Formula (101) in \cite{CoDo:Smatrix_curved}:
\bal{		
	\label{eq:_laHat12_CoordRep}
	\hat\la_{12}(\ta,\vc x)
	& = \Up(\ta)\,\et(\vc x)
		+ \coco{\Up(\ta)}\:\coco{\ki(\vc x)}.
	}
%

%
%
\subsubsection{Classical asymptotic field: Rod regions}
\label{sec:_Classical_Complex_ClassAsympt_Rods}
\noindent
For rod regions $M_R$, a real-valued solution $\la_{\del R}$ near its boundary $\Sibar_R$ consists of only one real-valued solution near $\Sibar_R$ which we denote by $\xi$ like the second solution above, that is: $\la_{\del R} = \xi$.
The complex structure on the boundary also has only one component here: $J_{\del R} = J_{\Sibar_R}=-J_{\Si_R}$, wherein the minus sign comes again from the opposite orientation of $\Sibar_R$.
Then, decomposition \eqref{eq:_Decomp_R+JI} can be written as
\bal{
	\la_{\del R} = \xi = \xi\htx R -J_{\Si_R}\xi\htx I.
	}
The classical asymptotic field \eqref{eq:_ClassAsymptoticField} is denoted here by
\bal{
	\hat\la_R \defeq \xi\htx R -\im\xi\htx I.
	}
Since for rod regions the boundary solution has only one component, the components $\xi\htx{R,I}_R$ cannot be calculated directly as in \eqref{eq:_laRI12}.
Rather, they are determined by the fact that the modes $X\htx a_{\om l m_l\!}(r)\, U_{\om l m_l}(t,\Om)$ are regular on the whole rod region, whereas the modes $X\htx b_{\om l m_l\!}(r)\, U_{\om l m_l}(t,\Om)$ may become singular.
That is, the real expansions \eqref{eq:modetub} of the components $\xi\htx{R,I}_R \in L\htx{int}_R$ must have $(\xi\htx R)\htx b_{\om l m_l}=(\xi\htx I)\htx b_{\om l m_l} \equiv 0$, and hence can be written like \eqref{eq:moderod}:
\bal{	
	\label{eq:_xiR_RealExp}
	\xi\htx R(t,r,\Om) 
	& = \int\!\!\xd\om\sum_{l,m_l} (\xi\htx R)\htx a_{\om l m_l\!}\,
		X\htx a_{\om l m_l\!}(r)\, U_{\om l m_l\!}(t,\Om),
		\\
	\label{eq:_xiI_RealExp}
	\xi\htx I(t,r,\Om) 
	& = \int\!\!\xd\om\sum_{l,m_l} (\xi\htx I)\htx a_{\om l m_l\!}\,
		X\htx a_{\om l m_l\!}(r)\, U_{\om l m_l\!}(t,\Om).
	}
In order to recover the real expansion of $\xi$, we act on \eqref{eq:_xiI_RealExp} with the complex structure \eqref{eq:_J_c_ab}, yielding
\bal{
	\label{eq:_xi_RealExp}
	\xi(t,r,\Om) 
	& = \int\!\!\xd\om\sum_{l,m_l} 
		\biiglrr{\xi\htx a_{\om l m_l\!}\,
					X\htx a_{\om l m_l\!}(r)\, U_{\om l m_l\!}(t,\Om)
					+\xi\htx b_{\om l m_l\!}\,
					X\htx b_{\om l m_l\!}(r)\, U_{\om l m_l\!}(t,\Om)
					},
		\\
	\label{eq:_xiab_Xi_xiRIab}
	\begin{pmatrix}
	\xi\htx a_{\om l m_l\!}\\
	\xi\htx b_{\om l m_l\!}
	\end{pmatrix}
	& = \;\Xi\;
	\begin{pmatrix}
	(\xi\htx R)\htx a_{\om l m_l\!}\\
	(\xi\htx I)\htx a_{\om l m_l\!}
	\end{pmatrix},
		\qquad\qquad
	\Xi =
	\begin{pmatrix}
	\;1\;\; & 
	\tfrac{-\Repart(\coco{c\htx a_{\om l m_l\!}}c\htx b_{\om l m_l\!})}
			{\Impart(\coco{c\htx a_{\om l m_l\!}}c\htx b_{\om l m_l\!})}\,
			\\
	\;0\;\; &
	\tfrac{-|c\htx b_{\om l m_l\!}|^2}
			{\Impart(\coco{c\htx a_{\om l m_l\!}}c\htx b_{\om l m_l\!})}\,
	\end{pmatrix}.
	}
Reading off the determinant of $\Xi$, we observe that it is nonvanishing precisely if $\Impart(\coco{c\htx a_{\om l m_l\!}}c\htx b_{\om l m_l\!}) \neq 0$, which is the usual condition that we require.
Hence $\Xi$ is invertible, and the inverse of \eqref{eq:_xiab_Xi_xiRIab} can be written as
\bal{
	\label{eq:_xiRIab_Xi_xiab}
	\begin{pmatrix}
	(\xi\htx R)\htx a_{\om l m_l\!}\\
	(\xi\htx I)\htx a_{\om l m_l\!}
	\end{pmatrix}
	& = \;\Xi^{-1}
	\begin{pmatrix}
	\xi\htx a_{\om l m_l\!}\\
	\xi\htx b_{\om l m_l\!}
	\end{pmatrix},
		\qquad\qquad
	\Xi^{-1} =
	\begin{pmatrix}
	\;1\;\; & 
	\tfrac{-\Repart(\coco{c\htx a_{\om l m_l\!}}c\htx b_{\om l m_l\!})}
			{|c\htx b_{\om l m_l\!}|^2}\,
			\\
	\;0\;\; &
	\tfrac{-\Impart(\coco{c\htx a_{\om l m_l\!}}c\htx b_{\om l m_l\!})}
			{|c\htx b_{\om l m_l\!}|^2}\,
	\end{pmatrix}.
	}
Given any complex structure \eqref{eq:_J_c_ab} through choosing $c\htx{a,b}_{\om l m_l\!}$, relation \eqref{eq:_xiRIab_Xi_xiab} with \eqref{eq:_xiR_RealExp} and \eqref{eq:_xiI_RealExp} provides the corresponding decomposition $\xi = \xi\htx R -J_{\Si_R}\xi\htx I$ of any given boundary solution $\xi$ as in \eqref{eq:_xi_RealExp}.
Now we can evaluate $\hat\la_R \defeq \xi\htx R -\im\xi\htx I$, resulting in the real expansion
\bal{
	\label{eq:_hat_la_R_RealExp}
	\hat\la_R(t,r,\Om) 
	& = \!\!\int\!\!\xd\om\!\sum_{l,m_l} 
			(\hat\la_R)\htx a_{\om l m_l\!}\,
			X\htx a_{\om l m_l\!}(r)\, U_{\om l m_l\!}(t,\Om),
	\qquad\quad
	(\hat\la_R)\htx a_{\om l m_l}\!
	= \xi\htx a_{\om l m_l\!}
  - \tfrac{c\htx a_{\om l m_l\!}}{c\htx b_{\om l m_l\!}}\,
	\xi\htx b_{\om l m_l\!}.
	}
Because of $c\htx{a,b}_{-\om, l,- m_l\!}=c\htx{a,b}_{\om l m_l\!}$
whereas $\xi\htx{a,b}_{-\om, l,- m_l\!}=\coco{\xi\htx{a,b}_{\om, l, m_l\!}}$, we get $(\hat\la_R)\htx a_{-\om,l,-m_l\!} \neq \coco{(\hat\la_R)\htx a_{\om l m_l\!}}$.
This confirms that $\hat\la_R$ is a not a real but a complex(ified) solution.
Using the transformation \eqref{eq:_rel_real_freq_momreps}, we can rewrite the real expansion \eqref{eq:_hat_la_R_RealExp} as a frequency expansion, yielding:
\bal{
	\label{eq:_hat_la_R_FreqExp}
	\hat\la_R(t,r,\Om) 
	& = \!\int\!\!\xd\om\!\sum_{l,m_l} 
		\biiglrr{(\hat\la_R)^+_{\om l m_l\!}
				\Up_{\om l m_l\!}(r)\, U_{\om l m_l\!}(t,\Om)
				+\coco{(\hat\la_R)^-_{\om l m_l\!}}\:
				\coco{\Up_{\om l m_l\!}(r)}\: \coco{U_{\om l m_l\!}(t,\Om)}
				},
		\\
	\label{eq:_laHatR-}
	\coco{(\hat\la_R)^-_{\om l m_l\!}}
	& = \tfrac1{2\im\,\Impart(\coco{c\htx a_{\om l m_l\!}}
											c\htx b_{\om l m_l\!})}
		\biiglrr{c\htx b_{\om l m_l}
				\coco{\xi\htx a_{\om l m_l}}
				-c\htx a_{\om l m_l}
				\coco{\xi\htx b_{\om l m_l}}
				}
		= \coco{\xi^-_{\om l m_l}},
		\\
	(\hat\la_R)^+_{\om l m_l\!}
	& = -\tfrac{\coco{c\htx b_{\om l m_l}}}{c\htx b_{\om l m_l\!}}\:
	\coco{(\hat\la_R)^-_{-\om,l,-m_l}}.
	}
As to be expected, in- and outgoing parts of $\hat\la_R$ are both determined by the solution $\xi$ near $\Sibar_R$.
However, while the outgoing part of $\hat\la_R$ is precisely the outgoing part of $\xi$, its ingoing part differs from the ingoing part of $\xi$.
Comparing \eqref{eq:_laHatR-} with \eqref{eq:_hat_la_R_RealExp},
we notice that we can write the rod's classical asymptotic field as
\bal{
	\label{eq:_hat_la_R_Ops}
	\hat\la_R(t,r,\Om) 
	& = \tfrac{X\htx a(r)}{c\htx b}\! \int\!\!\xd\om\!\sum_{l,m_l} 
	\tfrac{2\im\,\Impart(\coco{c\htx a_{\om l m_l\!}}
									c\htx b_{\om l m_l\!})}
			{2\im\,\Impart(\coco{c\htx a_{\om l m_l\!}}
									c\htx b_{\om l m_l\!})}\,
	\biiglrr{c\htx b_{\om l m_l} \xi\htx a_{\om l m_l\!}
				-c\htx a_{\om l m_l}	\xi\htx b_{\om l m_l\!}
				}\,
	U_{\om l m_l\!}(t,\Om),
		\notag
		\\
	& = \tfrac{X\htx a(r)\,2\im\,\Impart(\coco{c\htx a}c\htx b)}
					{c\htx b}\!
		\int\!\!\xd\om\!\sum_{l,m_l} \coco{\xi^-_{\om l m_l}}\:
		\coco{U_{\om l m_l\!}(t,\Om)}.
	}
Defining the following complex configurations as
\bal{		
	\coco{\ki(t,\Om)}
	& := \int\!\!\xd\om\!\sum_{l,m_l} \coco{\xi^-_{\om l m_l}}\:
			\coco{U_{\om l m_l}(t,\Om)},
	}
we find that the above $\hat\la_R$ coincides with the $\hat\la_R$
from Formula (114) in \cite{CoDo:Smatrix_curved}:
\bal{		
	\hat\la_R(t,r,\Om)
	& = \tfrac{X\htx a(r)\,2\im\,\Impart(\coco{c\htx a}c\htx b)}
					{c\htx b}\:
		\coco{\ki(t,\Om)}.
	}
%

%
\subsection{Real and complex inner products}
\label{sec:_Classical_R+C-InProd}
\noindent
The real g-product \eqref{eq:_def_g} induced by this complex structure can be written as
\bal{
	\label{eq:_g_+-}
	\txg_{\Si_\ta}\! \biglrr{\xi,\,\ze}
	:= 2 \om_{\Si_\ta}(\xi, J_{\Si_\ta} \ze)
	= -2\si \int\!\! \xd^3 k\; 
		\tilde w_{\vc k}(\ta) \mc W_{\vc k} (\ta)\,
		\Impart(\coco{c\htx a_{\vc k}} c\htx b_{\vc k})\,
		\biiglrr{\xi^+_{\vc k}\, \coco{\ze^-_{\vc k}}	
					+ \coco{\xi^-_{\vc k}}\,\ze^+_{\vc k}
					}.
	}
The positivity condition of the vacuum operator in Eq.~(64) of \cite{CoDo:Smatrix_curved} tells us that $-\si\,\Impart\biglrr{\coco{c\htx a_{\vc k}}c\htx b_{\vc k}}\,\mc W\lvc k (\ta) > 0$.
It is then quickly checked that in \eqref{eq:_g_+-} we have $\txg_{\Si_\ta}(\ph,\ph) > 0$ for real-valued $\ph \neq 0$, that is: our real g-product is positive definite.
Since both the symplectic and the complex structure change sign under orientation reversal, the real g-product is invariant under orientation reversal.
The complex inner product (\ref{eq:complex-inner-prod}) now becomes
\bal{\solinpro\xi\ze_{\Si_\ta}
	& := \txg_{\Si_\ta}\! \biglrr{\xi,\,\ze}
		+ 2\im\, \om_{\Si_\ta}\! \biglrr{\xi,\,\ze}
	\label{eq:_complex_inpro_+-}
	= -4\si \int\!\! \xd^3 k\; 
		\tilde w_{\vc k}(\ta)\mc W_{\vc k}(\ta)\,
		\Impart(\coco{c\htx a_{\vc k}} c\htx b_{\vc k})\;
		 \xi^+_{\vc k}\, \coco{\ze^-_{\vc k}}.
	}
Under orientation reversal we obtain the same expression for $\solinpro\xi\ze_{\Sibar_\ta}$, but with $\coco{\xi^-_{\vc k}}\,\ze^+_{\vc k}$.
Further, we can quickly check that the complex inner product is conjugate-linear in the first argument (with respect to $J_{\Si_\ta}$), and linear in the second :
\bal{
	\solinpro{(x\!+\! y J_{\Si_\ta})\xi}{\ze}_{\Si_\ta}
	& = (x\!-\!\im y) \cdot \solinpro{\xi}{\ze}_{\Si_\ta}\!,
		&
	\solinpro{\xi}{(x\!+\! y J_{\Si_\ta})\ze}_{\Si_\ta}
	& = (x\!+\!\im y) \cdot \solinpro{\xi}{\ze}_{\Si_\ta}.
	}
%
%
%
\section{Quantization}
\label{sec:_quant}
\noindent
In this section we briefly review the essential points two quantization schemes which implement the GBF for quantum fields. 
We glossing over many imporant details, for which we refer e.g.~to \cite{Oe:GBQFT,Oe:hol,Oe:aff,CoDo:Smatrix_curved}.
Several quantities outlined below are studied in more depth in Section \ref{sec:_Relation_SF_HOL}.
%
%
\subsection{Schr\"odinger-Feynman Quantization (SFQ)}
\label{sec:_quant_SF}
\noindent
In this scheme, quantum states of the field are described in the Schr\"odinger representation \cite{CoCoQu:SF, Hat, Jackiw} by wave functionals on spaces of field configurations, and amplitudes are calculated through a path integral quantization. 
The corresponding quantities are labeled by S (or by D when we switch to the Dirac/interaction picture).
The quantum state space $\cHS_{\Sigma}$ associated to each hypersurface $\Si$ consists of these wave functionals of field configurations on $\Si$.
The inner product of the Hilbert space $\cHS_{\Sigma}$ is formally given by
\bal{ 	
	\label{eq:inpro_schro}
	\inpro[\txS]{\al\htx S_\Si}{\be\htx S_\Si}
	\defeq \int_{C_\Si}\!\!\!\! \xD \vph \; 
	\coco{\al\htx S_\Si(\vph)} \, \be\htx S_\Si(\vph),
	}
where the integral is over the space $C_\Si$ of field configurations $\vph$ on the hypersurface $\Si$. 
(Field configurations do not depend on the orientation of $\Si$, and thus $C_\Sibar \equiv C_\Si$.)
The amplitude $\roS{M}$ for a boundary state $\psS{\del M}$ is defined heuristically as
\bal{		
	\label{eq:amplitude_general}
	\roS{M}(\psS{\del M}) = \int_{C_{\del M}}\!\! \xD \vph \; 
						\psS{\del M}(\vph) \, Z_M(\vph),
	}
wherein $Z_M$ is the field propagator encoding the field dynamics in the spacetime region $M$:
\bal{ 
	\label{eq:fieldprop}
	Z_M(\vph) 
	= \int_{\phi|_{\del M} = \vph} \!\!\!\! \xD \phi \; 
		\eu^{\im S_M(\phi)}.
	}
$S_M(\phi)$ is the action of the field in the region $M$, and the integration is extended over all field configurations $\phi$ (not only classical solutions) matching the boundary configuration $\vph$ on the boundary $\del M$. 
Next we consider the above objects for the different regions introduced in Section \ref{sec:_Classical}.

For interval (and tube) regions $\smash{M_{[\ta_1,\ta_2]}}$, the boundary hypersurface is the union of two disjoint hypersurfaces of constant $\ta$ each (hypercylinder surfaces of constant $r$ each), and hence the boundary state space $\cH_{\del[\ta_1,\ta_2]} = \cH_{\ta_1} \otimes \cH_{\ta_2}$ is the tensor product of the two boundary components' state spaces.
A state in this Hilbert space thus can be written as $\psi_{\ta_1} \otimes \smash{\coco{\psi_{\ta_2}}}$, wherein the complex conjugation of the second state is due to the opposite orientation of the second hypersurface (because both are oriented outwards).
The amplitude for this state takes the form
\bal{ 
	\label{eq:amplitude_general_interval}
	\roS{[\ta_1,\ta_2]} (\psS{\Si\ta_1} 
				\otimes \coco{\psS{\Si\ta_2}}\,) 
	& = \int_{C_1}\!\!\!\! \xD \vph^1 
		\int_{C_2}\!\!\!\!\xD \vph^2 \;
		\psS{\Si\ta_1}(\vph^1) \, \coco{\psS{\Si\ta_2}(\vph^2)} \;
		Z_{[\ta_1,\ta_2]}(\vph^1, \vph^2),
		\\
	\label{eq:propinterval}
	Z_{[\ta_1,\ta_2]}(\vph^1, \vph^2) 
	& = \int_{\begin{matrix}
					\scriptstyle	\phi|_{\Sigma_1}=\vph^1 \\ 
					\scriptstyle \phi|_{\Sigma_2}=\vph^2
					\end{matrix}
					} 
	\xD \phi \; \eu^{\im S_{[\ta_1,\ta_2]}(\phi)}.
	}
For tube regions $M_{[R_1,R_2]}$ the boundary state space is $\cH_{\del [R_1,R_2]} = \cH_{R_1} \otimes \cH_{R_2}$, and a state in this Hilbert space is $\psi_{R_1} \otimes \smash{\coco{\psi_{R_2}}}$.
This state's amplitude can be written just as \eqref{eq:amplitude_general_interval} with field propagator \eqref{eq:propinterval}, with $\ta_{1,2}$ replaced in both by $R_{1,2}$.

For a rod region $M_R$, the boundary state space is $\cH_{\del M_{R}} =  \cH_{R}$, and a state in this Hilbert space is $\coco{\psi_{R}}$ since $\Si_R$ is oriented inwards while $\del M_R =\Sibar_R$ is oriented outwards. 
This state's amplitude is
\bal{
	\label{eq:amplitude_general_rod}
	\roS{R} (\coco{\psS{\Si_R}}) 
	= \int_{C_R}\!\! \xD \vph^R \;
	\coco{\psS{\Si_R}(\vph^R)} \, Z_{R}(\vph^R),
	}
and the field propagator of the theory reads (with $S_{R}(\phi)$ the action of the rod region)
\bal{
	\label{eq:proprod}
	Z_{R}(\vph^R) = \int_{\phi|_{R} = \vph^R} \!\!\xD \phi \; 
							\eu^{\im S_{R}(\phi)}.
	}
Note that here the amplitude is calculated for a single state living on the boundary hypercylinder (and not for two states as usual).
This boundary state encodes both incoming and outgoing particles.
%
%
\subsection{Holomorphic Quantization (HQ)}
\label{sec:_Quant_Hol}
\noindent
We summarize here the Holomorphic Quantization as in Section 4 of \cite{Oe:hol} and Section 3 of \cite{Oe:aff}.
HQ arises as a particular kind of Geometric Quantization, see e.g.~Section 2 of \cite{Oe:aff} and Section 9.2 of \cite{Woo:geomquant}.
We recall the real and complex inner products $\txg_{\Si_\ta}(\cdot,\cdot)$ and $\{\cdot,\cdot\}_{\Si_\ta}$ on the space $L_{\Si_\ta}$ of solutions which are well defined in a neighborhood of $\Si_\ta$, see Section \ref{sec:_Classical_Structures}.
Quantum states $\psH{\Si_\ta}: L_{\Si_\ta} \to \complex$ in HQ are functionals similar to those of SFQ.
However, they are wave functions on spaces of classical solutions (representing phase space), whereas the SFQ wave functions live on configuration space.
The quantum state space $\cH\htx H_{\Si_\ta}$ of Holomorphic Quantization%
\footnote{As discussed in Section 3.3 of \cite{Oe:hol},
	states of HQ are actually functions on $\hat L_{\Si_\ta}$,
	which can be understood as an extended space of solutions, 
	including distributions. $\hat L_{\Si_\ta}$ is the algebraic dual 
	of the topological dual of $L_{\Si_\ta}$. Since the values of 
	a state $\psH{\Si_\ta}$ on $L_{\Si_\ta}$ completely fix 
	$\psH{\Si_\ta}$ on $\hat L_{\Si_\ta}$, we shall use the simplified
	notation of \cite{Oe:aff} and write just $L_{\Si_\ta}$.
	}
is the complex Hilbert space $\txH^2(L_{\Si_\ta},\nu_{\Si_\ta})$
of holomorphic square-integrable functions on $L_{\Si_\ta}$
with respect to the measure $\nu_{\Si_\ta}$ given by (32)
in \cite{Oe:aff}: $\xd\nu_{\Si_\ta}(\xi) =\xd\mu_{\Si_\ta}(\xi)\;
\exp\bglrr{-\tfrac12 \txg_{\Si_\ta}(\xi,\xi)}$.
Therein, $\mu_{\Si_\ta}$ is a (fictitious) Lebesgue measure on $L_{\Si_\ta}$ normalized to
$1 =\int_{\xi\in L_{\Si_\ta}}\!\!\!\! \xd\mu_{\Si_\ta}(\xi)\;
\exp\bglrr{-\tfrac12 \txg_{\Si_\ta}(\xi,\xi)}$,
making $\nu_{\Si_\ta}$ into a probability measure.
The inner product of the state space $\cH\htx H_{\Si_\ta}$ then can be written as
\bals{
	\inpro[\txH]{\al\htx H_{\Si_\ta}}{\be\htx H_{\Si_\ta}}
	=\int_{\xi\in L_{\Si_\ta}}\, \xd\nu_{\Si_\ta}(\xi)\;
	\coco{\al\htx H_{\Si_\ta}(\xi)}\, \be\htx H_{\Si_\ta}(\xi).
	}
The free amplitude for any holomorphic boundary state $\psH{\del M}$ is defined as the path integral
\bal{\label{eq:_DefAmpHol}
	\roH[0]{M}\biglrr{\psH{\del M}}
	=\int_{L\htx{int}_{\del M}}\!\! \xd\nu(\ph)\;
		\psH{\del M}(\ph).
	}
The measure $\xd\nu(\ph)$ is a probability measure like the one above, but on $L\,\!\htx{int}_{\del M}$.
Further, $\psH{\del M}$ is assumed to be integrable: $\psH{\del M} \in \mc L ^1(L\,\!\htx{int}_{\del M},\nu)$.
Holomorphic coherent states are integrable in this sense, and for them the integral \eqref{eq:_DefAmpHol} yields \eqref{eq:_FreeAmpHol_M}, see below.
%
%
\section{Relation between Schr{\"o}dinger-Feynman and Holomorphic
			Representation/Quantization}
\label{sec:_Relation_SF_HOL}
\noindent 
In this section we study the relation both on the level of the \emph{representation} of the quantum states (in the Schr\"odinger and in the Holomorphic Representation) and on the level of {quantization} (amplitudes in the Feynman and in the Holomorphic Quantization).
The relation between the Schr\"odinger representation and the holomorphic one has been explored in \cite{Oe:Sch-hol}, where a one-to-one correspondence between the two representations has been established. 
The basic idea is to view the two representations as particular polarizations of the prequantum Hilbert space defined in the framework of geometric quantization. 
Then, the inner product of a section in the Schr\"odinger Hilbert space with a section in the holomorphic Hilbert space allows the construction of a linear, isometric isomorphism $\mc B_{\Si_\ta}: \cH\htx S_{\Si_\ta} \to \cH\htx H_{\Si_\ta}$ between the two Hilbert spaces, which completely encodes the relation between the two representations, see Proposition 3.1 in \cite{Oe:Sch-hol}. 
In the following sections we discuss more details of this relation, and using the above mode expansions we calculate explicit expressions for all quantities involved in it, with the most important ones being the vacuum state, the coherent states and the free amplitudes.

The isomorphism $\mc B_{\Si_\ta}$ is an integral transform with the integral kernel $\tilde B_{\Si_\ta}: L_{\Si_\ta} \times Q_{\Si_\ta} \to \complex$,
\bal{
	\label{eq:_Corresp_ti_B}
	\tilde B_{\Si_\ta}(\xi,\vph) & = \exp\, \biiglrr{\!
			-\tfrac\im2 [j_{\Si_\ta}\!\vph, j_{\Si_\ta}\!\vph]
								_{\Si_\ta}
			\!\!-\! \tfrac12 \txg_{\Si_\ta}\! (j_{\Si_\ta}\!\vph, 
																j_{\Si_\ta}\!\vph)
			\!+\! \solinpro{j_{\Si_\ta}\!\vph}{\xi}_{\Si_\ta}
			\!\!-\! \tfrac12 \solinpro{j_{\Si_\ta}\!q_{\Si_\ta}\!\xi}
													{\xi}_{\Si_\ta}
			}.
	}
The first two terms in the exponential determine the 
Schr\"odinger vacuum state, while the last two terms determine the coherent states.
This is discussed in more detail in the following sections.
A holomorphic coherent state (see Section \ref{sec:_Correspond_Coherent}) is then mapped as follows to a Schr\"odinger coherent state, wherein $\vph = \ph|_{\Si_\ta}$:
\bal{
	\label{eq:_Corresp_Bbar}
	\psH[\xi]{\Si_\ta}(\ph)
	\mapsto \biglrr{\mc B_{\Si_\ta}^{-1}\psH[\xi]{\Si_\ta}} (\vph)
	= \coco{\tilde B_{\Si_\ta}(\xi,\vph)}.
	}
%
%
\subsection{Vacuum state} %
\label{sec:_Relation_SF_HOL_Vacuum}
\noindent
In this section we first review the vacuum in Schr\"odinger and holomorphic representation, and then study the correspondence between them.
The Schr\"odinger representation's vacuum is the Gaussian state $\psS[0]{\Si_\ta}$ derived in Equations (56) and (60) of \cite{CoDo:Smatrix_curved}, wherein $A_{\Si_\ta}$ is called the vacuum operator (for $\Up(\ta)$ see Section \ref{sec:_Classical_FrequencyRep}):

\bal{	
	\label{eq:_SchroVacGauss}
	\psSar[0]{\Si_\ta}\vph 
	& = \Norm[\text{S},0]{\Si_\ta}\;
	\exp\, \biiglrr{\!-\frac12  \int_{\Si_\ta}\!\! \xd^3 x\;
		\vph(\vc x)\,  \biglrr{A_{\Si_\ta}\vph}\! (\vc x)
		}	\;,
		\\
	\label{eq:_SchroVacOp}
	A_{\Si_\ta} 
	& = -\im \si\, 
	\sqrt{\abs{g^{(3)}g^{\ta\ta}}_{\Si_\ta}}\,
	\fracw{\coco{(\del_\ta\Up)(\ta)}}{\coco{\,\Up(\ta)\,}} \,,
		\\
	\label{eq:_N_VacSchro}
	\Norm[\text{S},0]{\Si_\ta} 
	& = \deth[1/4] 
		\biiglrr{\sqrt{\abs{g^{(3)}g^{\ta\ta}}_{\Si_\ta}}
					\tfracw{\abs{2\,\Impart (\coco{c\htx a} c\htx b)\,
									 \mc W (\ta)}}
								{2\pi\,\abs{\Up(\ta)}^2}
					},
	}
wherein the determinant for an operator $\mc O_{\Si_\ta}$ on $C_{\Si_\ta}$ is defined as 
\bal{	
	\deth[-1/2] \left( \tfrac{\mc O_{\Si_\ta}}{2\pi}\right)
	:= & \int_{C_{\Si_\ta}}\!\!\! \xD\vph\;
	\exp\,\biiglrr{-\frac12 \int_{\Si} \xd^3x\;
					\vph(\vc x)\, \biglrr{\mc O_{\Si_\ta}\vph}(\vc x)
						}.
	}
As discussed in Section \ref{sec:_Classical_ComplexStruct}, there are different choices of the functions $c\htx{a,b}_{\vc k}$ that lead to the same complex structure, and hence must lead to the same Schr\"odinger vacuum \eqref{eq:_SchroVacGauss}.
This can be verified easily by writing in the vacuum operator \eqref{eq:_SchroVacOp} as
\bal{
	\label{eq:_SchrodVac_ca1}
	\fracw{\coco{(\del_\ta\Up)(\ta)}}{\coco{\,\Up(\ta)\,}}
	= \fracw{\coco{c\htx a} \del_\ta\! X\htx a(\ta)
				+ \coco{c\htx b} \del_\ta\! X\htx b(\ta)}
				{\coco{c\htx a} X\htx a(\ta) + \coco{c\htx b} X\htx b(\ta)}
	= \fracw{\del_\ta\! X\htx a(\ta) + \tfracw{\coco{c\htx b}}
							{\coco{c\htx a}} \del_\ta\! X\htx b(\ta)}
				{X\htx a(\ta) + \tfracw{\coco{c\htx b}}
							{\coco{c\htx a}} X\htx b(\ta)},
	}
and taking $\tfracw{\coco{c\htx b}}{\coco{c\htx a}}$ as the new $\coco{c\htx b}$ which now completely determines the vacuum.
This corresponds to fixing $\coco{c\htx a} \equiv 1$.
In other words, multiplying any pair $(c\htx{a}_{\vc k}, c\htx{b}_{\vc k})$ 
by a complex function $f_{\vc k} \neq 0$ returns a new pair 
$(c\htx{a}_{\vc k}, c\htx{b}_{\vc k})$,
which induces the same vacuum state as the originals,
because $f_{\vc k}$ cancels in the quotient in \eqref{eq:_SchrodVac_ca1}.

In the holomorphic representation, the vacuum state is simply the constant wave function: $\psH[0]{\Si_\ta} \equiv 1$.
The correspondence formulas (44), (50) and (51) in \cite{Oe:Sch-hol} are a consequence of (\ref{eq:_Corresp_Bbar}), and map $\psH[0]{\Si_\ta} $ to the following Schr\"odinger state $\psS[0]{\Si_\ta}$, which is determined by the real bilinear form%
\footnote{Since the symplectic potential acts on two solutions,
	we would properly have to write $j_{\Si_\ta} \ki$,
	see (44) in \cite{Oe:Sch-hol}.
	However, the symplectic potential is sensitive only
	to the configuration of its second argument on $\Si_\ta$,
	which is just $\ki$.
	} 
$\Om_{\Si_\ta}$ on the space $C_{\Si_\ta}$ of field configurations on the hypersurface $\Si_\ta$ (for $\txg_{\Si_\ta}$ and $j_{\Si_\ta}$ see Section \ref{sec:_Classical_Structures}):
\bal{
	\psS[0]{\Si_\ta}(\vph)
	& = \Norm[\text{S},0]{\Si_\ta}\, 
		\exp \biglrr{-\tfrac12\,\Om_{\Si_\ta}(\vph,\vph)},
		\\
	\label{eq:_correspondence_50_Oe:Sch-hol}
	\Om_{\Si_\ta}(\vph,\ki)
	& = \txg_{\Si_\ta} (j_{\Si_\ta}\vph,	j_{\Si_\ta}\ki)
		 -\im\, [j_{\Si_\ta}\vph,\ki]_{\Si_\ta}.
	}
In the last line, we recognize (apart from a constant factor) the first two terms in the argument of the exponential in the right-hand-side of \eqref{eq:_Corresp_ti_B}, including the complex conjugation of \eqref{eq:_Corresp_Bbar}.
We show now, that this is precisely the Schr\"odinger vacuum \eqref{eq:_SchroVacGauss}.
In Sections \ref{sec:_Classical_SymplecPotStruct} - \ref{sec:_Classical_R+C-InProd} we have already calculated the necessary ingredients for evaluating \eqref{eq:_correspondence_50_Oe:Sch-hol}.
First we compute the real g-product (wherein $\vph,\ki$ are generic, real-valued configurations on $\Si_\ta$):
\bal{
	\txg_{\Si_\ta} (j_{\Si_\ta}\vph, j_{\Si_\ta}\ki)
	\,=\, 2\om_{\Si_\ta} (j_{\Si_\ta}\vph,
									J_{\Si_\ta} j_{\Si_\ta}\ki)\,
	= \tfracw{\im}2\si\! \int_{\Si_\ta}\!\!\!\!\xd^3 x\,
		\sqrt{|g^{(3)} g^{\ta\ta}|\,}\, \vph(\vc x)\,
		\biiglrr{\tfracw{(\del_\ta\Up)(\ta)}{\Up(\ta)}\! 
				\!-\tfracw{\coco{(\del_\ta\Up)(\ta)}}{\coco{\Up(\ta)}}
			}\,
		\ki(\vc x).
	}
Second, we compute the symplectic potential:
\bal{
	-\im\, [j_{\Si_\ta}\vph,\ki]_{\Si_\ta}
	& = \tfracw{\im}2\, \si\int_{\Si_\ta}\!\!\!\xd^3 x\, 
		\sqrt{|g^{(3)} g^{\ta\ta}|}\;\vph(\vc x)\,
		\biiglrr{-\tfracw{(\del_\ta\Up)(\ta)}{\Up(\ta)} 
				-\tfracw{\coco{(\del_\ta\Up)(\ta)}}{\coco{\Up(\ta)}}
			}
		\ki(\vc x).
	}
Summed as in \eqref{eq:_correspondence_50_Oe:Sch-hol}, they provide
\bal{
	\label{eq:_OmVacuumCoord}
	\Om_{\Si_\ta}(\vph,\ki)
	& = -\im\si\int_{\Si_\ta}\!\!\!\xd^3 x\,
		\sqrt{|g^{(3)} g^{\ta\ta}|}\; \vph(\vc x)\,
		\tfracw{\coco{(\del_\ta\Up)(\ta)}}{\coco{\Up(\ta)}}
		\ki(\vc x),
	}
which precisely reproduces the vacuum operator $A_{\Si_\ta}$ in \eqref{eq:_SchroVacOp} for $\vph=\ki$.
With the projectors \eqref{eq:_P+-}, it is then also quickly checked that (75) from \cite{Oe:Sch-hol} is fulfilled: $\Om_{\Si_\ta}\!(q_{\Si_\ta}\!P^+\xi, \vph) = -\im\,[P^+\xi, \vph]_{\Si_\ta}$, which encodes the one-to-one correspondence between Schr\"odinger vacuum states and complex structures.

Actually, the normalization factor $\Norm[\text{S},0]{\Si_\ta}$ is not included in the definition of the vacuum state denoted by $K\htx{S,0}_{\Si_\ta}$ in Formulas (44) and (51) in \cite{Oe:Sch-hol}, that is: $\psS[0]{\Si_\ta} = \Norm[\text{S},0]{\Si_\ta} K\htx{S,0}_{\Si_\ta}$.
This is due to different normalizations conventions: 
In \cite{Oe:Sch-hol}, $K\htx{S,0}_{\Si_\ta}$ is normalized with respect to the measure $\nu_Q$, see Section \ref{sec:_Quant_Hol}, whereas $\psS[0]{\Si_\ta}$ in \cite{CoDo:Smatrix_curved} is normalized with respect to the usual (fictitious, translation-invariant) measure $\mc D\vph$ of the Schr\"odinger representation. 
The measure $\mc D \vph$ comes with no fixed normalization, and therefore the Schr\"odinger vacuum $\psS[0]{\Si_\ta}$ includes a normalization constant giving it unit norm with respect to $\mc D \vph$.
The normalization factor then arises as follows, which precisely reproduces \eqref{eq:_N_VacSchro}:
\bal{
	\Norm[\text{S},0]{\Si_\ta}
	= \biiglrr{\int\!\! \xD\vph\;
					\exp -\txg_{\Si_\ta}(j_{\Si_\ta}\vph,j_{\Si_\ta}\vph)
					}^{-1/2}.
	}
As stated in \cite{Oe:Sch-hol}, there is one-to-one correspondence between the choice of the Schr\"odinger vacuum and the choice of complex structure.
Above we have seen that in our context both choices are determined by fixing the complex functions $c\htx{a,b}_{\vc k}$.
A \emph{unique} complex structure, and hence a unique Schr\"odinger vacuum, can be distinguished by applying additional conditions: 
For example, the energy condition in \cite{AshMag:QuantumCurved} equates the energy of a classical solution to that of the associated one-particle state, and thereby singles out one unique complex structure.
%
%
\subsection{Coherent states} 
\label{sec:_Correspond_Coherent}
\noindent
In this section we first review the coherent states in Schr\"odinger and holomorphic representations, and then study the correspondence between them.
For the Schr\"odinger representation (in the Dirac/interaction picture), we use the coherent state $\psD[\et]{\Si_\ta}$ defined in Equation (80) of \cite{CoDo:Smatrix_curved}, with $\et(\vc x)$ a complex configuration on $\Si_\ta$ called the state's characterizing function, 
\bal{	
	\label{zzz_SKG_SF_coherent_300}
	\psD[\et]{\Si_\ta}(\vph) 
	& = \Norm[\text{D},\et]{\Si_\ta}\;
	\exp \,\biiglrr{\int_{\Si_\ta}\!\!\!\xd^3 x\;
		 \vph(\vc x)\, \tfrac{w(\vc x)}{\coco{\Up(\ta)}}\, \et(\vc x)
			}\;
	\psS[0]{\Si_\ta}(\vph),
		\\
	\label{zzz_SKG_SF_coherent_310}
	\Norm[\text{D},\et]{\Si_\ta} 
	& = \exp\, \biiiglrr{\!-\!\tfrac12\!\int_{\Si_\ta}\!\!\!\xd^3 x\;
			\biiglrr{\et \tfracw{\Up(\ta)}{\coco{\Up(\ta)}} 
				w\mc K\htx D\, \et
				+  \et \,w\mc K\htx D\, \coco{\et}
				}(\vc x)
		}.
	}
Therein, $\mc K\htx D$ is the $\ta$-independent operator
\bal{
	\mc K\htx D
	& = \biglrr{-\si\, 2\Impart (\coco{c\htx a}c\htx b)\,
			\tilde w(\ta) \mc W (\ta)
			}^{-1}.
	}
In the holomorphic representation, coherent states are determined by the complex inner product.
We consider only their normalized version, given by
\bal{
	\label{Quantum_GBF_HQ_coherent_51}
	\psH[\xi]{\Si_\ta}(\ph) 
	&= \Norm[\txH,\xi]{\Si_\ta} \exp\,
		\biglrr{\tfrac12 \solinpro{\xi}{\ph}_{\Si_\ta} },
		&
	\Norm[\txH,\xi]{\Si_\ta} 
		&= \exp\,\biglrr{-\tfrac14 \txg_{\Si_\ta}(\xi,\xi) }.
	}
$\xi(\ta,\vc x)\in L_{\Si_\ta}$ is a solution near $\Si_\ta$ which we call the characterizing solution of the coherent state.
We assume that $\mc N^{\txH,\xi}_{\Si_\ta} \in \reals$, since $\xi$ is real-valued.
We observe that the Schr\"odinger coherent state \eqref{zzz_SKG_SF_coherent_300} is holomorphic in its characterizing function $\et(\vc x)$.
By contrast, the holomorphic one is antiholomorphic in $\xi(\ta,\vc x)$,
since the complex inner product is conjugate-linear in its first argument.
This is the reason for the complex conjugation in \eqref{eq:_Corresp_Bbar}.
The correspondence formulas (48), (49) and (52) in \cite{Oe:Sch-hol} are also consequences of \eqref{eq:_Corresp_Bbar}, and associate the following Schr\"odinger state to the holomorphic state $\psH[\xi]{\Si_\ta}(\ph)$:
\bal{
	\label{eq:_coherent_KS}
	\tilde K^{\txS,\xi}_{\Si_\ta} (\vph) 
	& = \psSar[0]{\Si_\ta}\vph
		\exp\, \biiglrr{\solinpro{\xi}{j_{\Si_\ta}\!\vph}_{\Si_\ta}
						-\tfrac12 \solinpro{\xi}
										{j_{\Si_\ta}\!q_{\Si_\ta}\!\xi}_{\Si_\ta}
								},
		\\
	& = \psSar[0]{\Si_\ta}\vph
		\exp\, \biiglrr{\Om_{\Si_\ta}\!(q_{\Si_\ta\!}\xi,\vph)
							+\!\im [\xi,\vph]_{\Si_\ta}\!
							-\!\tfrac12 \Om_{\Si_\ta}\!(q_{\Si_\ta\!}\xi,
																	q_{\Si_\ta\!}\xi)
							-\!\tfrac\im2 [\xi,\xi]_{\Si_\ta}\!
							}.
	}
In the first line, we recognize the (conjugated) last two terms of \eqref{eq:_Corresp_ti_B}.
The second line uses the bilinear vacuum form and the symplectic potential already calculated in \eqref{eq:_OmVacuumCoord} and \eqref{eq:_sympl_pot_tau_k}.
We now show the following correspondence:
\bal{
	\label{eq:_coherent_corresp}
	\tilde K^{\txS,\et\htx D_{12}}_{\Si_\ta} (\vph) 
	= \psD[\et]{\Si_\ta} (\vph),
	}
wherein $\et\htx D_{12}$ is the following real-valued solution, which arises from Equation (100) in \cite{CoDo:Smatrix_curved} by setting $\ki=\et$, and $\hat\et_{12}$ is from \eqref{eq:_laHat12_CoordRep} above:
\bal{\label{eq:_eta_D}
	\et\htx D_{12}(\ta,\vc x) 
	& = \mc K\htx D \hat\et_{12}(\ta,\vc x)
	= \mc K\htx D
		\biiglrr{\Up(\ta)\, \et (\vc x)
				+ \coco{\Up(\ta)}\: \coco{\et (\vc x)}
				}.
	}
In other words, the correspondence maps holomorphic coherent states to Schr\"odinger ones according to
\bal{
	\label{eq:_correspond_coher_HS}
	\mc B_{\Si_\ta}: \psD[\et]{\Si_\ta} 
	\mapsto \psH[\et\htx D_{12}]{\Si_\ta},
	}
wherein the characteristic solution and function are related through \eqref{eq:_eta_D}.
With \eqref{eq:_OmVacuumCoord} and \eqref{eq:_sympl_pot_tau_k} it is now  straightforward to calculate
\bal{
	\Om_{\Si_\ta}\! (q_{\Si_\ta}\!\et\htx D_{12}, \vph) 
	+ \im [\et\htx D_{12},\vph]_{\Si_\ta}
	& = \int_{\Si_\ta}\!\!\!\xd^3 x\;
		 \vph(\vc x)\, \tfrac{w(\vc x)}{\coco{\Up(\ta)}}\, \et(\vc x),
		\\
	-\tfrac12\Om_{\Si_\ta}\!(q_{\Si_\ta}\!\et\htx D_{12}, 
										q_{\Si_\ta}\!\et\htx D_{12}) 
	- \tfrac\im2 [\et\htx D_{12}, \et\htx D_{12}]_{\Si_\ta}
	& = -\frac12\!\int_{\Si_\ta}\!\!\!\xd^3 x\;
			\biiglrr{\et \tfracw{\Up(\ta)}{\coco{\Up(\ta)}} 
				w\mc K\htx D\, \et
				+  \et\, w\mc K\htx D\, \coco{\et}
				}(\vc x),
	}
This coincides exactly with \eqref{zzz_SKG_SF_coherent_300}, and hence confirms the identity \eqref{eq:_coherent_corresp}.
We end this section by comparing the inner products.
For the Schr\"odinger coherent states we found in (82) of \cite{CoDo:Smatrix_curved}:
\bal{		
	\label{zzz_SKG_SF_coherent_350}
	\inproo{\psD[\et]{\Si_\ta}}{\psD[\ki]{\Si_\ta}}
	& = \exp \int_{\Si_\ta}\!\!\!\xd^3 x\; 
		\biiglrr{ \coco\et\,w\KD \ki
			-\tfrac12 \coco\et\,w\KD \et
			- \tfrac12 \coco\ki \,w\KD \ki
			}(\vc x).
	}
For the inner product of normalized holomorphic coherent states we obtain from (35) and (36) in \cite{Oe:Sch-hol}:
\bal{
	\inpro{\psH[\ze]{\Si_\ta}}{\psH[\xi]{\Si_\ta}}
	& = \exp \biiglrr{\tfrac12\solinpro{\xi}{\ze}_{\Si_\ta}\!
						-\tfrac14\txg_{\Si_\ta}\!(\xi,\xi)
						-\tfrac14\txg_{\Si_\ta}\!(\ze,\ze)
						}.
	}
Using \eqref{eq:_g_+-} and \eqref{eq:_complex_inpro_+-}, it is straightforward to verify that $\inpro{\psH[\et\htx D_{12}]{\Si_\ta}}{\psH[\ki\htx D_{12}]{\Si_\ta}}$ exactly reproduces $	\inpro{\psD[\et]{\Si_\ta}}{\psD[\ki]{\Si_\ta}}$ in \eqref{zzz_SKG_SF_coherent_350}.
This confirms the unitarity of the isomorphism $\mc B_{\Si_\ta}$ stated in Proposition 3.1 of \cite{Oe:Sch-hol}.
%
%
\subsection{Free amplitudes} 
\noindent
In this section we verify that the free amplitudes of coherent states coincide for Schr\"odinger-Feynman (SFQ) and Holomorphic Quantization (HQ).
We start with the interval regions.
Evaluating the amplitude formula \eqref{eq:amplitude_general_interval} of SFQ for coherent states, in \cite{CoDo:Smatrix_curved} we have found:
\bal{
	\label{eq:_FreeAmplitSchrodInterval}
	\roS[0]{\intval\ta12} 
	\biiglrr{\psD[\et]{\Si\ta_1}\!\otimes\! 
				\coco{\psD[\ki]{\Si\ta_2}}\,}
	& = \exp \int\!\! \xd^3 x\;
		\biiglrr{ \et\,w\KD \coco\ki
			-\tfrac12 \coco\et\,w\KD \et 
			- \tfrac12 \coco\ki \,w\KD \ki
			}(\vc x).
	}
For HQ, we consider first a general region $M$.
Given $\la\in L_{\del M}$, the free amplitude for a normalized coherent state $\psH[\la]{\del M}\in\cHH_{\del M}$ has been calculated in (46) and (47) of \cite{Oe:hol}, using $\hat\la_M$ from \eqref{eq:_ClassAsymptoticField} above:
\bal{
 	\label{eq:_FreeAmpHol_M}
 	\roH[0]M \biglrr{\psH[\la]{\Si_\ta}}
	\,=\, 
	\mc N^{\txH,\la}_{\del M}\,
	\exp\left(\tfrac{1}{4}g_{\del M}(\hat\la_M,\hat\la_M)\right)
	\,=\,
	\exp\, \biiglrr{-\tfrac12 \txg_{\del M}(\la^\txI,\,\la^\txI)
							-\tfrac\im2 \txg_{\del M}(\la^\txR,\,\la^\txI)
							}.
	}
For an interval region, with $\la=(\ze,\xi)$ and
$\txg_{\del M} = \txg_{\Si{\ta_1}}\!\!+\! \txg_{\Sibar{\ta_2}} = 2\txg_{\Si{\ta_1}}$ this can be written as
\bal{
 	\label{eq:_FreeAmpHol_12}
 	\roH[0]{\intval\ta12}
	\biglrr{\psH[\ze]{\Si\ta_1} \otimes \coco{\psH[\xi]{\Si\ta_2}} }
	= \exp\, \biiglrr{- \txg_{\Si{\ta_1}}(\la^\txI_{12},\,\la^\txI_{12})
							-\im \txg_{\Si{\ta_1}}(\la^\txR_{12},\,\la^\txI_{12})
							}.
	}
With $\la\htx{R,I}_{12}$ from \eqref{eq:_laRI12} and \eqref{eq:_g_+-} it is quickly verified that $\roH[0]{\intval\ta12} (\psH[\et\htx D_{12}]{\Si\ta_1} \otimes \coco{\psH[\ki\htx D_{12}]{\Si\ta_2}} )$ yields the same expression as \eqref{eq:_FreeAmplitSchrodInterval}.
That is, on interval regions the amplitudes in Schr\"odinger-Feynman and Holomorphic Quantization coincide for coherent states related through the correspondence \eqref{eq:_correspond_coher_HS}.

Next, we consider rod regions. For SFQ, in \cite{CoDo:Smatrix_curved} 
we had evaluated the amplitude \eqref{eq:amplitude_general_rod} for coherent states:
\bal{		
	\label{eq:_FreeAmplitSchrodRod}
	\roS[0]{\Si_R}\biiglrr{\,\coco{\psD[\ki]{\Si_R} }\, } 
	& = \exp\, \biiglrr{-\frac12 \int_{\Si_R}\!\!\!\!
			\xd t\,  \xd\Om\; \biglrr{
					\coco\ki \tfracw{\coco{c\htx b}}{c\htx b}w\KD \coco\ki
					+ \coco{\ki}\, w\KD \ki
					}(t,\Om)
			}.
	}
For the rod region, with $\la=\xi$ and $\txg_{\del R} = \txg_{\Sibar_R}= \txg_{\Si_R}$, the holomorphic amplitude \eqref{eq:_FreeAmpHol_M} can be written as
\bal{
 	\label{eq:_FreeAmpHol_R}
 	\roH[0]R
	\biglrr{\coco{\psH[\xi]{\Si\ta_2}} }
	\,=\,
	\mc N^{\txH,\xi}_{\del M}\,
	\exp\left(\tfrac{1}{4}g_{\del M}(\hat\la_M,\hat\la_M)\right).
	}	
wherein $\xi\htx{R,I}_R$ are from \eqref{eq:_hat_la_R_FreqExp}.
With this and \eqref{eq:_g_+-} it is quickly verified that $\roH[0]R (\coco{\psH[\ki\htx D_{12}]{\Si\ta_2}} )$ yields the same expression as \eqref{eq:_FreeAmplitSchrodRod}.
That is, the SFQ and HQ amplitudes coincide also on rod regions for coherent states related through the correspondence \eqref{eq:_correspond_coher_HS}.
%
%
\section{Complex structure in phase space}
\label{sec:_ComplexStructPhase}
\noindent
In order to relate the vacuum wave functions in the Schr\"odinger representation and the complex structures defining the holomorphic one, it is useful to recall previous results that are unrelated to the GBF. 
In particular, the authors of \cite{CoCoQu:SF,CoCoQu:SF2} consider linear real scalar field theories in globally hyperbolic spacetime, which is generalized in Section 5.3 of \cite{Oe:Sch-hol}.
The theory is described by the symplectic vector space $(L_{\Si_\ta}, \om_{\Si_\ta})$.
The space of classical solutions $L_{\Si_\ta}$ (also called covariant phase space) is the canonical phase space (also called classical phase space) coordinatized by initial value data on a (backwards oriented) hypersurface $\Si_\ta$ consisting of field configurations $\vph(\vc x) = \ph(\ta,\vc x)|_{\Si_\ta}$ and the canonically conjugate momentum density $\pi = \sqrt{|g^{(3)}|} (\del_n \ph)|_{\Si_\ta} = \sqrt{|g^{(3)}g^{\ta\ta}|} (\del_\ta \ph)|_{\Si_\ta}$.
As above, $g^{(3)}$ is the determinant of the induced metric on $\Si_\ta$ and $\del_n$ is the normal derivative with respect to this hypersurface. 
In \cite{CoCoQu:SF,CoCoQu:SF2}, the hypersurface $\Si_\ta$ is taken to be a Cauchy surface.
Here, we do not require this: We only require that the space $L_{\Si_\ta}$ of classical solutions 
(that are well defined and bounded near $\Si_\ta$) indeed can be coordinatized by the above initial data on $\Si_\ta$.
Writing a solution $\ph$ as $\tbinom{\vph}{\pi}$, the symplectic structure \eqref{eq:_sympl_struct_tau_x} reads as
%
\bal{
	\om_{\Si_\ta}\biglrr{\tbinom{\vph_1}{\pi_1}, \tbinom{\vph_2}{\pi_2}}
	& = -\tfrac\si2 \int_{\Si_\ta}\!\!\! \xd^3 x\;
		\biglrr{\vph_1 \pi_2 - \vph_2 \pi_1}(\vc x).
}
The most general form of the complex structure $J_{\Si_\ta}$ is expressed in \cite{CoCoQu:SF} through four real-valued, linear, continuous operators $A,B,D,C$ as (the $A$ here \emph{is not} the vacuum operator denoted by $A_{\Si_\ta}$)
\bal{ 
	\label{eq:_J_PhaseSpace}
	J_{\Si_\ta} \binom\vph\pi 
	= \begin{pmatrix}
		A & B \\
		D & C
		\end{pmatrix}
		\binom\vph\pi .
	}
In order to get $J^2_{\Si_\ta} = -\One$, the operators must satisfy 
$A^2 + BD = C^2 + DB = -\One$ and $AB + BC = DA + CD = 0$.
In \cite{CoCoQu:SF,CoCoQu:SF2}, the vacuum state in the Schr\"odinger representation is expressed in terms of these operators:
\bal{
	\label{eq:_SchrodVac_CoCoQu}
	\psS[0]{\Si_\ta}(\vph) 
	= \exp \left( -\frac{\si}{2} \int_{\Si_\ta}\!\!\! \xd^3 x\;
	 		\vph(\vc x) \biglrr{(B^{-1}\!\! -\! \im \,C B^{-1})\vph}(\vc x) 
			\right).
	}
We included the overall sign $\si$, which is positive in \cite{CoCoQu:SF,CoCoQu:SF2} while in general we have $\si:=\sign (g_{00}g^{\ta\ta})$, see Section \ref{sec:_Classical_SymplecPotStruct}.
For \eqref{eq:_SchrodVac_CoCoQu} to be well defined, $B$ needs to be invertible.
By comparing $(B^{-1}\!\! -\! \im \,C B^{-1})$ with the vacuum operator $A_{\Si_\ta}$ in \eqref{eq:_SchroVacOp} above, we can read off that $B^{-1}$ must be the real part $\AR_{\Si_\ta}$ of the vacuum operator given in (63) of \cite{CoDo:Smatrix_curved} as the invertible operator
(due to $\Up(\ta)\neq 0$ and $\Impart (\coco{c\htx a} c\htx b) \neq 0$, see Section IV.B in \cite{CoDo:Smatrix_curved})
%
\bal{ 
	B^{-1} = \si\AR_{\Si_\ta} 
	& = - \,\sqrt{\abs{g^{(3)}g^{\ta\ta}}_{\Si_\ta}}\,
		\fracw{\Impart (\coco{c\htx a} c\htx b)\,\mc W (\ta)}
				{\abs{\Up(\ta)}^2}.
	}
We can also read off that $-C B^{-1}$ must be the imaginary part of the vacuum operator, yielding
\bal{
	-C	& 
	= \frac{\Repart\biglrr{\coco\Up\,\del_\ta\!\Up}(\ta)}
			{\mc W(\ta)\, \Impart(\coco{c\htx a} c\htx b)}.
	}
By similar arguments as given in \cite{CoDo:Smatrix_curved} for $\Up(\ta)\neq 0$, we also have $\del_\ta\!\Up(\ta)\neq 0$.
Hence $\Repart\biglrr{\coco\Up\,\del_\ta\!\Up}(\ta) \neq 0$, which makes $C$ invertible. 
Moreover, since we assume $\Up(\ta)$, $c\htx{a,b}$ and $X\htx{a,b}(\ta)$ to commute with each other, we have also $B$ and $C$ commuting.
Then, $AB+BC = 0$ implies that $A = -C$.
Finally, $C^2 + DB = -\One$ implies that
\bal{
	D & = \frac{\sqrt{\abs{g^{(3)}g^{\ta\ta}}_{\Si_\ta}}\;
						|\del_\ta\!\Up(\ta)|^2}
					{\mc W(\ta)\, \Impart(\coco{c\htx a} c\htx b)}.
	}
Hence $A,B,D,C$ are all well defined and invertible here and also commute with each other.
In order to relate the complex structure in phase space \eqref{eq:_J_PhaseSpace} to the one in solution space \eqref{eq:_J_+-},
we rewrite the relation \eqref{recover_freqrep_interval} between initial data $\tbinom{\vph(\vc x)}{\pi(\vc x)}$ and associated solution $\ph(\ta,\vc x)$:
\renewcommand\arraystretch{1.5}
\bal{
	\label{recover_freqrep_interval_phase}
	\begin{pmatrix}
		\ph^+\lvc k \\
		\coco{\ph^-_{-\vc k}}
	\end{pmatrix}
	& = \int_{\Si_\ta}\!\! \xd^3 x \;
		w_{\vc k}(\vc x)\, \coco{U\lvc k (\vc x)}\;\,F\,
		\begin{pmatrix}
			\vph(\vc x) \\ \pi(\vc x)
		\end{pmatrix},
		\\
	F & = \fracw1{2\im\,\Impart(\coco{c\htx a} c\htx b)\, \mc W(\ta)}
		\begin{pmatrix}
		\!-\coco{(\del_\ta\! \Up)(\ta)} 
		& \;\;\;\coco{\Up(\ta)}/\sqrt{\abs{g^{(3)}g^{\ta\ta}}_{\Si_\ta}}
			\\
		\;\;(\del_\ta\! \Up)(\ta) & 
		\;-\Up(\ta)/\sqrt{\abs{g^{(3)}g^{\ta\ta}}_{\Si_\ta}}\,
		\end{pmatrix}.
	}
The action of the complex structure \eqref{eq:_J_PhaseSpace} then can be written as
\bal{
	\label{recover_freqrep_interval_J_phase}
	\begin{pmatrix}
		(J_{\Si_\ta}\ph)^+\lvc k \\
		\coco{(J_{\Si_\ta}\ph)^-_{-\vc k}}
	\end{pmatrix}
	& = \int_{\Si_\ta}\!\! \xd^3 x \;
		w_{\vc k}(\vc x)\, \coco{U\lvc k (\vc x)}\;\,F\,
		\begin{pmatrix}
			A & B \\
			D & C
		\end{pmatrix}
		\begin{pmatrix}
			\vph(\vc x) \\ \pi(\vc x)
		\end{pmatrix}.
	}
With the above expressions for $A,B,D,C$, it is now easy to check that
\bal{
		F\;
		\begin{pmatrix}
			A & B \\
			D & C
		\end{pmatrix}
	= \begin{pmatrix}
			-\im & 0 \\
			\;\;0 & \im
		\end{pmatrix}
		\; F
	\qquad \implies \qquad
	\begin{pmatrix}
		(J_{\Si_\ta}\ph)^+\lvc k \\
		\coco{(J_{\Si_\ta}\ph)^-_{-\vc k}}
	\end{pmatrix}
	= 	\begin{pmatrix}
			-\im\, \ph^+\lvc k \\
			\;\im\,\coco{\ph^-_{-\vc k}}\,
		\end{pmatrix},
	}
\renewcommand\arraystretch{1.0}%
which precisely reproduces the complex structure \eqref{eq:_J_+-}.
That is, the two complex structures \eqref{eq:_J_c_ab} and \eqref{eq:_J_+-} we have found on solution space and \eqref{eq:_J_PhaseSpace} on phase space induce each other.

For consistency, let us verify whether the momentum and configuration subspaces $M_{\Si_\ta}$ and $N_{\Si_\ta}$ of solution space are generated by \eqref{recover_freqrep_interval_phase} from the $\pi$-subspace and $\vph$-subspace of phase space.
Expanding $\pi(\vc x)$ with \eqref{eq:modeinterval}, we find that inserting $\tbinom{0}{\pi(\vc x)}$ into \eqref{recover_freqrep_interval_phase} indeed yields a solution $\ph(\ta,\vc x)$ fulfilling \eqref{eq:_ph+_ph-_UpUp}, that is: $\ph \in M_{\Si_\ta}$, while $\smash{\tbinom{\vph(\vc x)}{0}}$ yields a solution $\ph(\ta,\vc x)$ fulfilling \eqref{eq:_ph+_ph-_DotUpDotUp}, that is: $\ph \in N_{\Si_\ta}$.
Therefore, on phase space we can write the associated projectors \eqref{eq:_PM_freq} simply as
\bal{
	P_M \begin{pmatrix} \vph \\ \pi \end{pmatrix}
	& = \begin{pmatrix} 0 & 0 \\ 0 & 1 \end{pmatrix}
		\begin{pmatrix} \vph \\ \pi \end{pmatrix}
	= \begin{pmatrix} 0 \\ \pi \end{pmatrix},
		&
	P_N \begin{pmatrix} \vph \\ \pi \end{pmatrix}
	& = \begin{pmatrix} 1 & 0 \\ 0 & 0 \end{pmatrix}
		\begin{pmatrix} \vph \\ \pi \end{pmatrix}
	= \begin{pmatrix} \vph \\ 0 \end{pmatrix}.
	}
With this and the complex structure \eqref{eq:_J_PhaseSpace} it is then clear that Eq.~(126) of \cite{Oe:Sch-hol} holds:
\bal{\label{eq:_ABDC_MN_PJ}
	\begin{split}
	A &: N_{\Si_\ta} \to N_{\Si_\ta}
		\qquad\qquad
	\,A = P_N J_{\Si_\ta},
		\\
	B &: M_{\Si_\ta} \to N_{\Si_\ta}
		\qquad\qquad
	B = P_N J_{\Si_\ta},
		\\
	D &: N_{\Si_\ta} \to M_{\Si_\ta}
		\qquad\qquad
	D = P_M J_{\Si_\ta},
		\\
	C &: M_{\Si_\ta} \to M_{\Si_\ta}
		\qquad\qquad
	\!C = P_M J_{\Si_\ta}.
	\end{split}
	}
That is, via \eqref{recover_freqrep_interval_phase} we find that $\tbinom{A\vph}{0}$ and also $\tbinom{B\pi}{0}$ induce $\ph \in N_{\Si_\ta}$, whereas $\tbinom{0}{D\vph}$ and also $\tbinom{0}{C\pi}$ induce $\ph \in M_{\Si_\ta}$.
The only property of $A,B,D,C$ necessary for this to hold is that they have real eigenvalues fulfilling $A_{\vc k}=A_{-\vc k}$, ditto for $B,D,C$ (plus the correct placement of the factor $\sqrt{|g^{(3)}g^{\ta\ta}|}$ converting scalar fields into scalar densities and vice versa).
Hence what makes $A,B,D,C$ special is not so much that they fulfill \eqref{eq:_ABDC_MN_PJ}, but rather that together they constitute a complex structure on phase space which via \eqref{recover_freqrep_interval_phase} induces the ''real'' complex structure \eqref{eq:_J_c_ab} and the ''frequency'' complex structure \eqref{eq:_J_+-}.
%
\section{Summary}
\label{sec:_Conclusions}
\noindent
The purpose of the present work is to confirm and develop the results presented in \cite{Oe:Sch-hol} about the correspondence between the Schr\"odinger and holomorphic representations used in QFT within the GBF. 
(For quantum states on spacelike hypersurfaces, this correspondence is independent of the GBF, 
 which comes into play here only for hypersurfaces that are not spacelike.)
We study the case of Klein-Gordon theories in a wide class of curved spacetimes. 
The novelty of our approach is to construct this correspondence by working (formally) 
with mode expansions of solutions of the equation of motion. 
In particular, here we have provided explicit expressions for the relevant structures 
used in the two representations and thereby confirmed their correspondence. 
To be more precise, we parametrize the relation between complex structures on solution space 
and Schr\"odinger vacua through two complex functions $c\htx a_{\vc k}, c\htx b_{\vc k}$ on momentum space,
and find which choices of these functions induce the same complex structure, and hence the same vacuum state.
We have also calculated the (generalized transition) amplitudes of coherent states and shown their equality:
For the Schr\"odinger representation this is done via Schr\"odinger-Feynman Quantization (SFQ),
and for the holomorphic representation via Holomorphic Quantization (HQ).
Finally, we also calculate explicitly the complex structure on phase space, 
which relates our results with previous ones obtained in literature.
The complete agreement of the free amplitudes calculated in the rigorous Holomorphic Quantization
with those coming from the Schr\"odinger-Feynman Quantization justifies
the heuristic procedures of the latter.
Future work will focus on extending this correspondence to interactions,
which can already be treated in SFQ \cite{CoDo:Smatrix_curved}, but not in HQ.
%
%
\begin{acknowledgments}
\noindent
It is a pleasure to thank Robert Oeckl (CCM-UNAM Morelia) for many stimulating discussions.
This work was supported in part by UNAM-DGAPA-PAPIIT project grant IA105416 (DC), as well as 
CONACyT scholarship 213531 (MD), UNAM-DGAPA-PAPIIT project grant IN100212 (MD) 
and PRODEP scholarship DSA/103.5/16/4816 of the SEP of Mexico (MD).
\end{acknowledgments}
%
%

%
%
\end{document}